\documentclass[12pt]{article}
\usepackage{fix-cm}
\pdfoutput=1
\usepackage[top=30truemm,bottom=30truemm,left=25truemm,right=25truemm]{geometry}
\usepackage{authblk}
\usepackage{amsthm}
\usepackage{amsmath,amssymb,bm}
\usepackage{cite}
\usepackage[]{hyperref}
\usepackage[titletoc]{appendix}
\usepackage{url}
\usepackage{graphicx,color}
\usepackage[dvipsnames]{xcolor}
\usepackage{braket}
\usepackage{qcircuit}
\usepackage{here}
\usepackage{mathrsfs}
\usepackage{bbm}
\usepackage{titlesec}

\numberwithin{equation}{section}

\hypersetup{
    colorlinks = true,
    allcolors = {Fuchsia}
}

\newtheorem*{definition*}{Definition}
\newtheorem*{assumption*}{Assumption}

\def\sinh{{\mathrm{sinh}}}
\def\cosh{{\mathrm{cosh}}}
\def\tanh{{\mathrm{tanh}}}

 \def\d{{\delta}}

 \def\a{{\alpha}}

 \def\CA{{\cal A}}

 \def\CM{{\cal M}}

 \def\r{\rho}

\def\be{\begin{equation}}
\def\ee{\end{equation}}
\def\ba{\begin{eqnarray}}
\def\ea{\end{eqnarray}}

\def\nn{\nonumber \\}

\def\ib{{\bar{\imath}}}
\def\jb{{\bar{\jmath}}}
\def\id{{\mathbbm{1}}}
\def\hb{\bar{h}}
\def\d{\mathrm{d}}
\def\i{\mathrm{i}}

\usepackage{setspace} 

\AtBeginDocument{%
  {\footnotesize\global\footnotesep=0.8\baselineskip}%
}

\begin{document}

\begin{titlepage}
\thispagestyle{empty}

\begin{flushright}
\end{flushright}

\bigskip

\begin{center}
\noindent{\bf \Large Wormholes with Ends of the World
}\\
\vspace{1.4cm}

{\bf Diandian Wang$^I$, Zhencheng Wang$^i$, and Zixia Wei$^I$}
\vspace{1cm}\\

{\it
$^I$Jefferson Physical Laboratory, 
Harvard University, Cambridge, MA 02138, USA
}\\[1.5mm]
{\it $^{i}$Department of Physics, University of Illinois Urbana-Champaign, Urbana, IL 61801, USA}\\[1.5mm]

\vskip 3em
\end{center}

\begin{abstract}

We study classical wormhole solutions in 3D gravity with end-of-the-world (EOW) branes, conical defects, kinks, and punctures. These solutions compute statistical averages of an ensemble of boundary conformal field theories (BCFTs) related to universal asymptotics of OPE data extracted from the 2D conformal bootstrap. Conical defects connect BCFT bulk operators; branes join BCFT boundary intervals with identical boundary conditions; kinks (1D defects along branes) link BCFT boundary operators; and punctures (0D defects) are endpoints where conical defects terminate on branes. We provide evidence for a correspondence between the gravity theory and the ensemble. In particular, the agreement of the $g$-function dependence results from an underlying topological aspect of the on-shell EOW brane action, from which a BCFT analog of the Schlenker-Witten theorem also follows.

\end{abstract}

\end{titlepage}

\newpage
\setcounter{page}{1}

\begingroup
    \fontsize{11}{12}\selectfont
    \tableofcontents
\endgroup

\newpage
\section{Introduction}
\label{sec:intro}

The instanton method offers a powerful approach for studying non-perturbative effects in quantum theories. Owing to its semiclassical nature, it is especially valuable when the full structure of the underlying quantum theory remains elusive. In the path integral formulation of quantum gravity, Euclidean wormholes are a class of instantons and contribute non-perturbatively in $1/G_N$. Notably, contributions from so-called replica wormholes \cite{AHMST19,PSSY19} have provided key insights into the black hole information paradox \cite{Hawking75,Penington19,AEMM19,AMMZ19} by reproducing the Page curve \cite{Page93}.

At the same time, the existence of Euclidean wormhole contributions also sharpens a previously underappreciated aspect of the AdS/CFT correspondence \cite{Maldacena97}. In the path integral formulation of AdS/CFT \cite{GKP98,Witten98}, the partition function of a CFT defined on a $d$-dimensional manifold $\Sigma$, denoted as $Z_{\rm CFT}[\Sigma]$, is dual to a gravitational path integral over all $(d+1)$-dimensional asymptotically AdS spacetimes whose asymptotic boundary is given by $\Sigma$. From the CFT point of view, when $\Sigma$ consists of two connected components $\Sigma_1$ and $\Sigma_2$, the partition function factorizes as $Z_{\rm CFT}[\Sigma_1\sqcup \Sigma_2]=Z_{\rm CFT}[\Sigma_1]Z_{\rm CFT}[\Sigma_2]$. However, this seems inconsistent with the gravity result if wormholes connecting $\Sigma_1$ and $\Sigma_2$ are included in the gravitational path integral \cite{MM04,AOP07}. This is often referred to as the factorization puzzle. Similar phenomena occur in asymptotically flat spacetimes \cite{Coleman88,GS88,GS87}. The existence of wormholes as stable saddles and their ubiquity \cite{MM04,MS21} make the puzzle even more pronounced. 

One modern understanding of the factorization puzzle is that certain simple gravitational path integrals in AdS may not correspond to quantities in a single CFT, but rather to an ensemble of CFT data. This is particularly precise in lower-dimensional models of gravity. For example, it has been found that the path integral of the Jackiw-Teitelboim gravity in AdS$_2$ corresponds to a matrix integral \cite{SSS19} (see also \cite{MM20} for a 2D topological gravity model). Further investigations of this idea led to the discovery of a duality between wormhole contributions in AdS$_3$ Einstein gravity coupled to massive particles and statistical moments of an ensemble of CFT$_2$'s with large central charge \cite{CCHM22}. The pursuit of a complete understanding of AdS${}_3$ Einstein gravity as an ensemble of quantum theories remains an active and evolving area of research. Recent progress has been made across several fronts, including on-shell configurations \cite{BdB20,BdBL21,BBJNS23,CEZ23,dBLPS23,CEZ24,JRW24,dBLP24,CJ21}, off-shell contributions \cite{CJ20,MT20,Yan23,Yan25,BDFPR25}, and geometries with conical defects \cite{BCM20,CCHM22,PSV24}.

It is worth emphasizing that in 3D Einstein gravity (with a negative cosmological constant) coupled to conical defects, the ensemble interpretation yields a quantitatively precise correspondence \cite{CCHM22} between semiclassical gravity and the universal asymptotics of CFTs \cite{CGMP18,Kusuki18,CMMT19,ABdBL21}.  Wormhole solutions also exist in higher-dimensional gravity theories, but higher-dimensional CFTs lack the infinite-dimensional Virasoro symmetry group, which is essential in tightly constraining the dynamics and correlation functions in 2D CFTs. Therefore, while it is important to try to obtain a more quantitative understanding of the story of ensemble averaging in higher dimensions, we will stay in three bulk dimensions and try to test the rigidity of this quantitative duality.

In this work, we construct a class of wormhole solutions in Euclidean AdS$_3$ Einstein gravity with massive particles and dynamical end-of-the-world (EOW) branes \cite{KR00,KR01}. Such wormhole configurations connect multiple asymptotic boundaries, which themselves are 2D manifolds with borders.\footnote{To avoid confusion, we will often refer to the boundaries of such 2D surfaces as ``borders" to distinguish them from boundaries of 3D manifolds.} Due to this feature, it is natural to expect a connection with 2D boundary conformal field theories (BCFTs) \cite{Cardy89}, i.e.,~2D CFTs defined on bordered Riemann surfaces.\footnote{We will often think of BCFTs as extensions of CFTs rather than analogs, so the term ``BCFT'' refers to a CFT on a general orientable Riemann surface, whether compact or bordered.} Indeed, we will demonstrate using simple examples that the contributions coming from these wormholes nicely match the results from averaging over BCFT data in an ensemble that is consistent with universal asymptotics obtained from BCFT bootstrap (which includes the CFT bootstrap as a subset) \cite{CMMT19,Kusuki18,Kusuki21,NT22}. This largely extends the paradigm of the correspondence between the wormhole contributions in semiclassical gravity and the ensemble average over non-gravitational quantum theories. We will also present no-go theorems forbidding certain wormhole configurations with EOW branes, implying that certain data in the BCFT do not exhibit averaging.

When the asymptotic boundary $\Sigma$ has only one connected component, the correspondence between AdS with EOW branes and a BCFT living on $\Sigma$ is known as the AdS/BCFT correspondence, first proposed by Takayanagi \cite{Takayanagi11,FTT11}. While being a seemingly over-simplified bottom-up model, the AdS$_3$/BCFT$_2$ correspondence works surprisingly well even when non-trivial gravitational interactions between the EOW branes and massive particles are accounted for, and it is able to reproduce many results of 2D conformal bootstrap in a single BCFT setting \cite{KW22}. From this point of view, our work significantly extends the validity of the AdS/BCFT correspondence to a more general setting involving more than one BCFT. Note that a case where a specific set of boundary OPE data is averaged was previously considered in \cite{Kusuki22}, but that does not originate from wormhole contributions and is different from the ensemble averaging we consider in this paper.

The subjects of BCFT and AdS/BCFT themselves are also rich topics. BCFTs play an important role in the theory of open strings \cite{Polchinski98}, in the study of critical many-body systems with boundaries \cite{AL91,OA96}, in modeling quench dynamics \cite{CC05,CC06}, and in providing a rigorous framework for entanglement entropy in CFT \cite{OT14,KMOP23,RLS25,KOP25}. The AdS/BCFT correspondence has been used to model thermalization dynamics in holography \cite{HM13,Ugajin13,STW18,CNSTW19,AKSTW20}, to study evaporating black holes via the so-called double holography \cite{AMMZ19,RSvRWW19,CFHMR19,SvRW20,CMNRS20,GK20}, to probe cosmological dynamics on the EOW brane \cite{CRSvRW19,AS19,AB24,FKKT25}, and to model measurement and measurement-induced phase transitions in holography \cite{NSTW16,ABCSJ22,AGJS23,KKSTTW23}.

\paragraph{Summary of results} Here we provide an outline of the structure of the paper, as well as a summary of the main results. We start by reviewing some basics and setting up the problem: 
\begin{itemize}
    \item In Section~\ref{sec:BCFTensemble}, we review the basics of 2D BCFT, define a BCFT ensemble at large $c$, and explain its relation to the universal asymptotics obtained from conformal bootstrap.
    
    Our BCFT ensemble consists of the following data:
    \begin{equation}
        \{c, g_a; (h_i,\bar{h}_i),h_I; C_{ijk},B_{IJK}^{(abc)},D_{iI}^{(a)} \}    
    \end{equation}
    with fixed central charge $c\gg 1$, a list of $g$-functions, one for each boundary condition $a$; dimensions of bulk operators $(h_i,\bar h_i)$ and dimensions of boundary operators $h_I$; and random OPE coefficients $C_{ijk}$ (bulk-bulk-bulk), $B_{IJK}^{(abc)}$ (boundary-boundary-boundary), and $D_{iI}^{(a)}$ (bulk-boundary). There is a discrete spectrum of both bulk and boundary operators with dimensions below the black hole threshold, and a continuum spectrum above it. We specify the first and second moments of the ensemble such that they are consistent with the universal asymptotics \cite{CMMT19,Kusuki18,NT22,Kusuki21}. 
    \item Section~\ref{sec:3dgrav} introduces the 3D gravity model we study in this paper: Einstein gravity in AdS$_3$ coupled to conical defects, EOW branes, one-dimensional defects along branes (which we call kinks), and zero-dimensional defects at the intersection of conical defects and branes (which we call punctures). In this model, conical defects correspond to local BCFT bulk operators below the black hole threshold, EOW branes connect borders with the same boundary conditions, and kinks correspond to local BCFT boundary operators below the black hole threshold. As we will see, the existence of punctures is important for consistency with the BCFT ensemble.
    
    We specify the gravitational action of this model. While there is no \textit{a priori} choice for this action, we will adopt a somewhat minimalist one, which, perhaps a little surprisingly, turns out to reproduce exact quantitative details of the ensemble. 
\end{itemize}
In the next four sections, we focus on the construction of wormhole solutions in this model and their relation to the BCFT universal asymptotics:
\begin{itemize}
    \item To begin with, in Section~\ref{sec:g}, we discuss the holographic calculation of the $g$-function, and its relation to the EOW brane tension $T$. We then present both a conceptual argument and an explicit calculation to demonstrate that there are no wormhole solutions connecting two empty disks (i.e., disks with no conical defects and kinks). This gives a holographic demonstration of the fact that $g$ is not an ensemble-averaged quantity.
    \item Wormhole solutions of pure gravity supported by conical defects, as found in \cite{CCHM22}, remain valid solutions in our model, and they reproduce the universal asymptotics of $C_{ijk}$. 
    
    In our model, there are many more wormholes. In Section~\ref{sec:tensionless}, we present our first set of new wormhole solutions: wormholes supported by conical defects, punctures, and \emph{tensionless} EOW branes. These can be constructed as $\mathbb{Z}_2$ quotients of wormhole solutions without EOW branes. We provide examples demonstrating how they contribute to ensemble-averaged observables.
    \item To construct wormholes with non-zero EOW brane tension, we outline the general procedure of gluing or removing a ``wedge" to or from the tensionless wormholes identified earlier. In Section~\ref{sec:tension}, we present this construction in the special case where there are EOW branes, conical defects, punctures, but no kinks. In this case, the action of such a wedge turns out to be ``topological", in the sense that it only depends on the brane tension and the brane topology. In the case with no kinks, the topological nature of the action reproduces the $g$-dependence of certain BCFT OPE statistics and multi-copy observables.
    \item We finally arrive at the most general class of wormhole solutions in Section~\ref{sec:gen}, namely, those involving conical defects, tensionful branes, kinks, and punctures. We show that the previously introduced ``wedge trick" continues to work. Regarding the action of the wedge, we conjecture a topological expression and test it in a simple example. If this conjecture holds generally, it would reproduce the expected $g$-function dependence for the most general ensemble-averaged BCFT observables.
\end{itemize}
Finally, we end with various comments on miscellaneous topics: 
\begin{itemize}
 
    \item In Section \ref{sec:remarks}, we begin by discussing the role of non-vanishing first moments and the agreement between gravity and the universal asymptotic expressions for higher moments. We then apply the results we derived to obtain a BCFT version of the Schlenker-Witten theorem. We also remark on some similarities between states above and below the threshold using the perspective of topological quantum field theory (TQFT).
    \item In Section~\ref{sec:disc}, we briefly summarize our findings and outline potential directions for future research.
\end{itemize}

\section{BCFT review and ensemble}
\label{sec:BCFTensemble}

In this section, we start by presenting the preliminaries of 2D BCFTs in Section~\ref{sec:BCFT_preliminaries}. In the process, we will introduce conventions and notations used in this paper. After that, we will write down an ensemble average over BCFT data and explain its relation with the BCFT universal asymptotics found in \cite{Kusuki18,CMMT19,ABdBL21,Kusuki21,NT22}. This ensemble averaging will be used to compare with the wormholes constructed later in this paper. 
Experts of BCFT may skip this part and go directly to Section~\ref{ssec:ensemble}. Finally, we demonstrate in Section~\ref{ssec:obs} how the ensemble gives non-factorized answers for two-copy observables and more generally multi-copy ones, along the lines of \cite{CCHM22}. 

\subsection{Preliminaries}
\label{sec:BCFT_preliminaries}
\begin{figure}
    \centering
    \includegraphics[width=\linewidth]{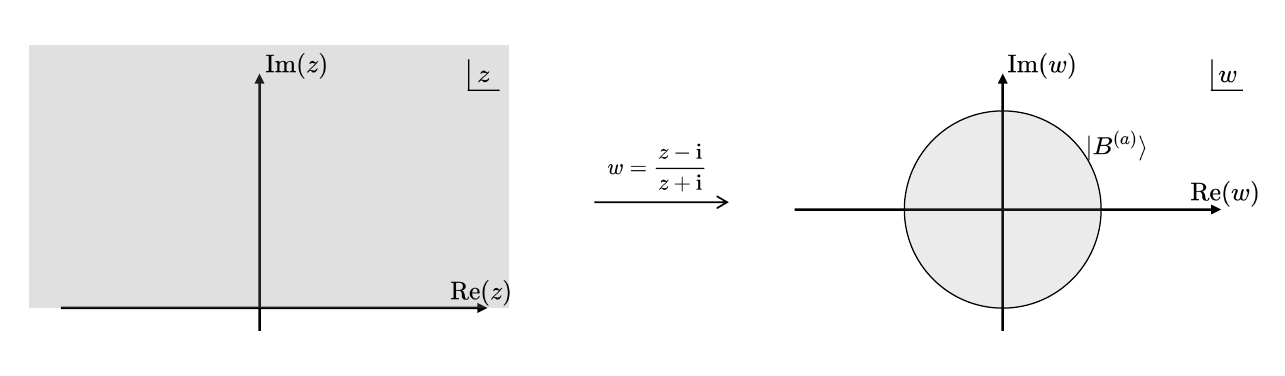}
    \caption{The upper half plane (shaded gray in the left panel) is a simple example of a 2D bordered surface. Under conformal transformation, it can be mapped to a unit disk (shaded gray in the right panel). From radial quantization, the partition function of such a disk can be regarded as the inner product $\langle B^{(a)}| \id \rangle$, for some boundary state $|B^{(a)}\rangle$ living on the circular border.}
    \label{fig:bcft}
\end{figure}

Let us start by considering a CFT on a bordered 2D surface. One of the simplest examples is the upper half plane (UHP), which is the ${\rm Im}(z) \geq 0 $ part of the complex plane $\mathbb{C}$ parameterized by $(z,\bar{z}) = (x+\i\tau, x-\i\tau)$ and possesses one boundary. See the left panel of Figure~\ref{fig:bcft} for a sketch. The conformal symmetries of a 2D CFT on $\mathbb{C}$ are represented by two copies of Virasoro symmetry, which will be partially broken when defined on the UHP. If one imposes the boundary condition 
\begin{align}\label{eq:boundary_condition}
    \left.\left(T(z) - \bar{T}(\bar{z})\right)\right|_{\rm bdy} = 0,
\end{align}
then the conformal symmetries are maximally preserved and turn out to be one copy of Virasoro symmetry. A CFT defined on a manifold with borders, where each boundary condition maximally preserves the conformal symmetries, is called a BCFT. 
A conventional 2D CFT is specified by three sets of data: the central charge $c$, the primary operators $\phi_i$ with conformal dimensions $(h_i, \bar{h}_i)$, and the OPE coefficients $C_{ijk}$ appearing in the three-point functions of primary operators on $\mathbb{C}$:
\begin{align}
    \braket{\phi_i(z_1)\phi_j(z_2)\phi_k(z_3)}_{\mathbb{C}} = C_{ijk}~ z_{12}^{h_k-h_i-h_j}\bar{z}_{12}^{\bar{h}_k-\bar{h}_i-\bar{h}_j} z_{23}^{h_i-h_j-h_k}\bar{z}_{23}^{\bar{h}_i-\bar{h}_j-\bar{h}_k} z_{31}^{h_j-h_k-h_i}\bar{z}_{31}^{\bar{h}_j-\bar{h}_k-\bar{h}_i}, 
\end{align}
where $z_{12} \equiv z_1 - z_2$, etc. 
These data are also inherited by the BCFT. The primary operators and the OPE coefficients inherited from the parent CFT are called bulk primaries and bulk-bulk-bulk OPE coefficients, respectively. 

However, specifying a BCFT requires additional data, as there are now boundary primary operators living on the borders themselves. Let us use $\phi_I^{(ab)}$ to denote a boundary primary operator labeled by $I$, whose one side is a boundary condition labeled by $a$, and the other side is a boundary condition labeled by $b$. 
In particular, a boundary primary with $b\neq a$ is called a boundary-condition-changing operator.
Throughout this paper, we use lowercase letters $a,b,c, ...$ to label boundary conditions, lowercase letters $i,j,k,...$ to label the bulk primaries, and capital letters $I,J,K,...$ to label boundary primaries. 

The new BCFT data involving the boundary condition and the boundary primaries are encoded in the following objects.

\paragraph{Disk partition function}
Performing the conformal transformation
\begin{align}
    w = \frac{z-\i}{z+\i},
\end{align}
the UHP plane parameterized by $(z,\bar{z})$ is mapped to a unit disk parameterized by $(w, \bar{w})$. 
See the right panel of Figure~\ref{fig:bcft}. 
This disk partition function is also called the $g$-function, 
\begin{align}
    g_a \equiv Z_{\rm disk}^{(a)},
\end{align}
which is an important characterization of the boundary condition labeled by $a$. 

\paragraph{Bulk-boundary two-point function}
Consider the UHP with a conformal boundary condition labeled by $a$. The correlation function between a bulk primary operator $\phi_i(z)$ and a boundary primary operator $\phi_I^{(aa)}(x)$ takes the form 
\begin{align}
    \braket{\phi_i(z)\phi_I^{(aa)}(x)}_{\rm UHP} = D_{iI}^{(a)} ~ (2{\rm \,Im}(z))^{h_I-h_i-\bar{h}_i} |z-x|^{\bar{h}_i-{h}_i-h_I} |\bar{z}-x|^{{h}_i-\bar{h}_i-h_I}.
\end{align}
The coefficient $D^{(a)}_{iI}$ appearing here is called the bulk-boundary OPE coefficient, which is a new set of data in BCFT. Note that we have chosen the normalization such that
\begin{align}
    D^{(a)}_{\id\id} = g_a. 
\end{align}
\paragraph{Boundary-boundary-boundary three point function}
Another set of data of BCFT comes from the three-point function of boundary primaries:  
\begin{align}
    \braket{\phi_I^{(ab)}(x_1) \phi_J^{(bc)}(x_2) \phi_K^{(bc)}(x_3)}_{\rm UHP} = B_{IJK}^{(abc)}~x_{12}^{h_K - h_I - h_J} x_{23}^{h_I - h_J - h_K} x_{31}^{h_J - h_K - h_I}, 
\end{align}
where $B^{(abc)}_{IJK}$ appearing here are called the boundary-boundary-boundary OPE coefficients. We have fixed the normalization such that
\begin{align}
    B^{(aaa)}_{\id\id\id} = g_a,\quad B^{(aba)}_{IJ\id} = \sqrt{g_ag_b}\,\delta_{IJ},
\end{align}
and the one-point function $B^{(aaa)}_{I\id\id}$ vanishes for any non-identity $\phi_I$.

\paragraph{}
We have now introduced all the data needed to specify a BCFT: the central charge $c$, the bulk primaries $\phi_i$ with the conformal dimensions $(h_i,\bar{h}_i)$ and the OPE coefficients $C_{ijk}$ between them, the $g$-function $g_a$ for each boundary condition labeled by $a$, the boundary primaries $\phi_I^{(ab)}$ with the conformal dimension $h_I$ and the OPE coefficients $B_{IJK}^{(abc)}$ between them, and the bulk-boundary OPE coefficients $D^{(a)}_{iI}$.

\begin{figure}
    \centering
\includegraphics[width=.9\linewidth]{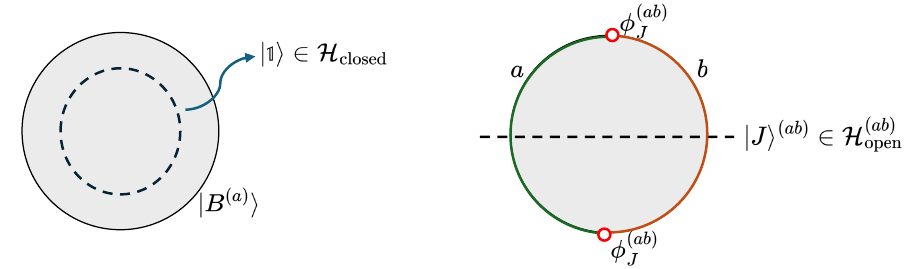}
    \caption{Closed Hilbert space (left) and open Hilbert space (right).}
    \label{fig:Hopenclosed}
\end{figure}

It is convenient to look at the BCFT data through the lens of the state-operator correspondence. 
Let us again consider the unit disk sketched in Figure~\ref{fig:bcft}. 
From the radial quantization point of view, the disk partition function can be regarded as an inner product between the CFT vacuum state $\ket{\id}$ and a conformal boundary state $\ket{B^{(a)}}$, where $a$ labels the boundary condition imposed at $|w|^2 =1$. This disk partition function can therefore be written as
\begin{align}
    g_a \equiv Z_{\rm disk}^{(a)} = \braket{B^{(a)}|\id} .
\end{align}
The Hilbert space obtained from the radial quantization on the disk consists of states defined on a circle. Therefore, we call it the closed-string Hilbert space $\mathcal{H}_{\rm closed}$, or closed Hilbert space for short. The state-operator correspondence associated with the radial quantization on the disk relates bulk operators of BCFT to states in the closed Hilbert space $\mathcal{H}_{\rm closed}$. See the left panel of  Figure~\ref{fig:Hopenclosed}.

Consider now how to write a boundary state $\ket{B^{(a)}}$ as a linear combination of primary states and descendant states in the closed Hilbert space $\mathcal{H}_{\rm closed}$. 
The constraint \eqref{eq:boundary_condition} is equivalent to 
\begin{align}\label{eq:bc_eq}
    (L_n - \bar{L}_{-n}) \ket{B^{(a)}} = 0,\quad (n \in \mathbb{N}). 
\end{align}
A set of states satisfying this equation can be constructed in the following way. Consider a scalar bulk primary $\phi_j$ with $h_j = \bar{h}_j$, 
and the Verma module associated with its holomorphic part. 
Let $\ket{j;N}$ be a descendant state labeled by $N$ in this Verma module. It is known that
\begin{align}
    |j\rangle\rangle \equiv \sum_N \ket{j;N}\otimes U\overline{\ket{j;N}} , 
\end{align}
which is often called the Ishibashi state \cite{Ishibashi88}, serves as a solution of \eqref{eq:bc_eq}. Here $\overline{\ket{j;N}}$ is the anti-holomorphic counterpart of ${\ket{j;N}}$, and $U$ is an antiunitary operator. A true boundary state can be written as a linear combination of the Ishibashi states, 
\begin{align}
    \ket{B^{(a)}} = \sum_j D^{(a)}_{j\id} |j\rangle\rangle,
\end{align}
where $D^{(a)}_{j\id}$ are constrained by bootstrapping the annulus partition function \cite{Cardy89,CMW21}.  

Besides the radial quantization, we can also consider the quantization along the ${\rm Im}(w)$ direction. In this case, the open-string Hilbert space (or open Hilbert space for short) $\mathcal{H}_{\rm open}^{(ab)}$ naturally appears, consisting of states defined on an interval with boundary condition $a$ imposed on the one side and $b$ imposed on the other. See the right panel of Figure \ref{fig:Hopenclosed}. A boundary primary $\phi_J^{(ab)}$ is related to a primary state $\ket{J}^{(ab)}$ in $\mathcal{H}_{\rm open}^{(ab)}$ via the state-operator correspondence.

\paragraph{Summary of conventions and notations}\label{sec:BCFT_notations}
As a summary, a BCFT can be specified with the following set of data:
\begin{itemize}
    \item $c$: the central charge;
    \item $(h_i,\bar{h}_i)$: spectrum of bulk operators, also known as closed-string states;
    \item $g_a$: the $g$-function of each boundary condition;
    \item $h_I$: spectrum of boundary operators, also known as open-string states;
    \item $C_{ijk}$: bulk-bulk-bulk OPE, or three-point function of bulk operators on the sphere or the plane;
    \item $B_{IJK}^{(abc)}$: boundary-boundary-boundary OPE, or three-point function of boundary operators on the disk (or the UHP);
    \item $D_{iI}^{(a)}$: bulk-boundary OPE, or two-point function of one bulk operator and one boundary operator on the disk (or the UHP) with boundary condition $a$.  
\end{itemize}
The indices $a,b,\dots$ label boundary conditions, one for each boundary interval between two boundary operators, and the normalization convention is such that\footnote{Note that this convention is the same as that used in \cite{CL91,NT22} but different from that of \cite{Kusuki21,Kusuki22,KW22}.}
\begin{align}
    g_a = D^{(a)}_{\id\id} = B^{(aaa)}_{\id\id\id}.
\end{align}

\subsection{Ensemble and universal asymptotics}
\label{ssec:ensemble}
We would like to define an ensemble of BCFTs and explain its relation with the BCFT universal asymptotics studied in \cite{Kusuki18,CMMT19,ABdBL21,Kusuki21,NT22}. This BCFT ensemble will be compared with the wormhole solutions constructed later in this paper.

We will use the Liouville parameters $Q,b,P$ to express the central charge and conformal weights: 
\begin{align}
    Q=b+b^{-1},~~~ c=1+6 Q^2,~~~h=\frac{Q^2}{4}+P^2, 
\end{align}
which are particularly convenient for expressing universal OPE asymptotics.

Similar to the CFT ensemble defined in \cite{CCHM22}, we fix the central charge $c\gg 1$ and assume that the BCFT ensemble consists of the following spectrum of {primary} operators:
\begin{itemize}
    \item A unique normalizable vacuum state (i.e. the state corresponding to the identity operator) in the closed Hilbert space $\mathcal{H}_{\rm closed}$;
    \item A unique normalizable vacuum state (i.e. the state corresponding to the identity operator) in the open Hilbert space $\mathcal{H}_{\rm open}^{(aa)}$ when two boundaries have the same boundary condition (note that no vacuum state can exist when two boundaries have different boundary conditions, simply because the identity operator acts trivially on the boundary condition);
    \item A finite, discrete list of scalar bulk operators with dimensions below the black hole threshold with $\frac{c-1}{32}<h_i=\bar{h}_i<\frac{c-1}{24}$;
    \item A finite, discrete list of boundary operators also with dimensions below the black hole threshold with $\frac{c-1}{32}<h_I<\frac{c-1}{24}$; 
    \item A continuous spectrum of states with Cardy density $\rho^{(\text{closed})}(P,\bar{P})=\rho_0(P) \rho_0(\bar{P})$ in the closed Hilbert space above the black hole threshold:  $h_i, \bar{h}_i \geq \frac{c-1}{24}$, {where $\rho_0(P)=4 \sqrt{2}\, \sinh (2 \pi P / b) \,\sinh (2 \pi P b)$};
    \item A continuous spectrum of states with Cardy density $\rho_{ab}^{(\text{open})}(P) = g_ag_b \rho_0(P)$ in the open Hilbert space $\mathcal{H}^{(ab)}_{\rm open}$ above the black hole threshold: $h_I \geq \frac{c-1}{24}$.\footnote{While the black hole threshold for bulk operators is well known, that for boundary operators may be less familiar. Our definition here follows \cite{Kusuki22,KW22}, which is when the AdS geometry dual to the boundary primary state transits from being a portion of the conical defect geometry to a portion of the BTZ black hole \cite{Kusuki22,KW22}.}
\end{itemize}
We have placed lower bounds $(h_i=\bar h_i>\frac{c-1}{32}, h_I>\frac{c-1}{32})$ on the conformal dimensions of the discrete lists of sub-threshold bulk and boundary operators so that any two such operators would combine to form a black hole. Consequently, we do not need to consider sub-threshold multi-twist operators built from them.

For the discrete list of bulk operators, we will also use an alternative parameterization for their conformal weights, following the convention of \cite{CCHM22}:
\begin{align}\label{eq:etadim}
    h_i=\bar{h}_i=\frac{c}{6} \eta_i(1-\eta_i), \qquad \eta_i<\frac{1}{2}.
\end{align}

We defined the BCFT ensemble by treating the OPE coefficients for primary operators as random ensembles with moments given by universal formulae obtained from bootstrap. In particular, one important expression is the universal OPE function \cite{CMMT19}
\begin{align}\label{eq:C0def}
    C_0(P_1, P_2, P_3)=\frac{\Gamma_b(2 Q) \Gamma_b(\frac{Q}{2} \pm \i P_1 \pm \i P_2 \pm \i P_3)}{\sqrt{2}\, \Gamma_b(Q)^3 \prod_{k=1}^3 \Gamma_b(Q \pm 2 \i P_k)},
\end{align}
where $\Gamma_b(Q)$ is the double Gamma function, and the expression should be read as taking the product over all choices of the $\pm$ signs. To simplify notations, we will often replace $P_i$ by just $i$ (and $\bar{P}_i$ by $\ib$) and use the notation $|\cdot|^2$ to denote multiplication of a function by its anti-holomorphic counterpart. In other words, 
\begin{equation}
    |C_0(ijk)|^2\equiv C_0(ijk)C_0(\ib \jb \bar{k})\equiv C_0(P_i, P_j, P_k) C_0(\bar{P}_i, \bar{P}_j, \bar{P}_k).
\end{equation}

Let us now specify the first moments and the second moments of the bulk OPE coefficients for non-identity operators $\phi_i$, $\phi_j$, and $\phi_k$:
\begin{align}
    \overline{C_{i j k}} &= 0, \label{eq:C_first}\\
    \overline{C_{i j k}C^*_{lmn}}&=|C_0(ijk)|^2[\delta_{i l} \delta_{j m} \delta_{k n}+\delta_{i m} \delta_{j n} \delta_{k l}+\delta_{i n} \delta_{j l} \delta_{k m}\nn
    &+(-1)^{\ell_i+\ell_j+\ell_k}( \delta_{i n} \delta_{j m} \delta_{k l}+\delta_{i m} \delta_{j l} \delta_{k n}+ \delta_{i l} \delta_{j n} \delta_{k m})]. \label{eq:C_second}
\end{align}
When one, two, or three of the indices are set to the identity $\id$, $C_{ijk}$ becomes a two-point function, a one-point function, or the sphere partition function, respectively, all of which contain no degrees of freedom. Furthermore, to make the mapping to 3D gravity precise, the second equation should hold if at least one of $i,j,k$ is above the black hole threshold, or
if 
\begin{align}
    \eta_i+\eta_j+\eta_k > 1,
\end{align}
and is zero otherwise. These are the same as the CFT ensemble defined in \cite{CCHM22}. As we will discuss further in Section~\ref{ssec:firstmom}, the vanishing of the first moment of $C_{ijk}$ is a choice to make the 3D gravity model simple. A quick explanation is that, at least in the case when all three operators are scalar operators with $\frac{c-1}{32} < h_i< \frac{c-1}{24}$, which correspond to conical defects in gravity, a non-vanishing $\overline{C_{ijk}}$ would require for example the existence of a junction joining three conical defects. Those are not part of the model of \cite{CCHM22}, and our model inherits this condition. The reason for the choice of the second moment of $C_{ijk}$ \eqref{eq:C_second} is less arbitrary and more non-trivial. As we will see, it is obtained by considering two-boundary wormholes in AdS$_3$, and the exact same expression appears in another context, namely in 2D conformal bootstrap. It was found in \cite{CMMT19}, by applying the fusion matrix method to the analytic bootstrap \cite{Kusuki18,CGMP18,KM19}, that \eqref{eq:C_second} is the universal expression for averaged heavy operators, where the averaging comes from coarse-graining over states in a single CFT. 

The ensemble discussed above was for the bulk OPE coefficients. Let us now move on to considering the averaging of the additional data required for BCFTs. To begin with, we would like to specify the first moments and the second moments of boundary OPE coefficients $B^{(abc)}_{IJK}$ for non-identity operators $\phi_I^{(ab)}$, $\phi_J^{(bc)}$, and $\phi_K^{(ac)}$ as
\begin{align}
\overline{B_{IJK}^{(abc)}}&=0,\label{eq:B_first}\\
\overline{B_{IJK}^{(abc)}B^{*(def)}_{LMN}}&=C_0(IJK) \label{eq:B_second}
\\
& \times(\delta_{IL} \delta_{JM} \delta_{KN}\delta_{ad}
\delta_{be}
\delta_{cf}
+\delta_{IM} \delta_{JN} \delta_{KL}
\delta_{ae}
\delta_{bf}
\delta_{cd}
+\delta_{IN} \delta_{JL} \delta_{KM}
\delta_{af}
\delta_{bd}
\delta_{ce}).\nonumber 
\end{align}
These expressions are closely resembling the expressions for $C_{ijk}$, and we will understand why in Section~\ref{sec:tensionless}. Notice that there is no spin associated with boundary operators because they only have one conformal weight corresponding to the existence of only one Virasoro symmetry algebra. However, unlike $C_{ijk}$, the $B_{IJK}$ with two indices swapped are independent, which is why the expression above contains only three terms, not six.

For the bulk-boundary OPE coefficients $D^{(a)}_{i\id}$, the situation is somewhat more intriguing. By considering the bootstrap of the BCFT partition function on the annulus, it was shown in \cite{Kusuki22} that setting $\overline{D^{(a)}_{i\id}}=0$ would disallow the discrete list of below-the-threshold boundary-condition-changing operators we specified as part of our ensemble.\footnote{Removing such operators produces a simpler but less non-trivial ensemble. We will comment on it further in Section~\ref{ssec:firstmom}.}
As we are interested in a more interesting and non-trivial correspondence between a 3D gravity theory and a BCFT ensemble, we do want to include these sub-threshold boundary operators, so we choose 
\begin{align}
    \overline{D^{(a)}_{i\id}}\ne0 .
\end{align}
More precisely, for non-identity $i$ and $I$, we specify the first and second moments to be
\begin{align}
\overline{D^{(a)}_{iI}}&=0,\label{eq:D_first}\\
\overline{D^{(a)}_{iI}D^{(b)}_{jJ}}&=  C_0(i\ib I)\delta_{ij}\delta_{IJ}\delta_{ab}, \label{eq:D_second}
\end{align}
where we have used $D^{(a)}_{iI}=D^{*(a)}_{iI}$, but for non-identity $i$ and identity $I=\id$, we choose
\begin{align}
\overline{D^{(a)}_{i\id}}&\ne 0,\label{eq:Diidave}\\
\overline{D^{(a)}_{i\id}D^{(b)}_{j\id}}|_{\text{conn.}}&=
    \begin{cases}
        \rho_0(P_i)^{-1}\delta_{ij}\delta_{ab},&h_i>\frac{c-1}{24},
        \\
        0,&h_i<\frac{c-1}{24}.
    \end{cases}  \label{eq:Diid2}
\end{align}

So far, we have presented the first and the second moments for each type of data in the ensemble. (There are no mixed second moments.) We would like to proceed without specifying the higher moments at this point, but we will comment on them when they become relevant. 

In Section~\ref{sec:3dgrav}, we will construct a gravity model and demonstrate that it reproduces all the statistical moments specified above and more.

\subsection{Ensemble observables}\label{ssec:obs}

Now that we have an ensemble, if we take two copies of the same quantity, it does not necessarily factorize. In \cite{CCHM22}, many examples of two-copy observables were worked out explicitly, including $n$-point functions on the sphere, one-point functions on the torus, and $0$-point functions on the genus-two Riemann surface. We will now demonstrate that the BCFT ensemble works similarly through an example.

Consider one bulk operator $\phi_i$ and two boundary operators $\phi^{(ab)}_I$ and $\phi^{(ab)}_J$ on a disk with boundary conditions $a$ and $b$ on the two boundary intervals separated by the two insertions:
\begin{align}\label{fig:1bulk2bdy}
\vcenter{\hbox{\includegraphics[height=3cm]{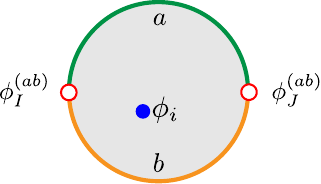}}} ~,
\end{align}
which we denote by $G_{iIJ}$. Now, consider the two-copy observable
\begin{align}\label{eq:2copyeg1}
G_{iIJ}(m_1) G_{iIJ}^{\prime}(m_2)=\langle \phi_i(z_1)\phi_I^{(ab)}(x_1)\phi_J^{(ab)}(x_2)\rangle_{\rm UHP}
\langle
 \phi_i(z_1')\phi_I^{(ab)}(x_1')\phi_J^{(ab)}(x_2')
 \rangle_{\rm UHP},
\end{align}
with real moduli parameters $m_1$ and $m_2$. Let us expand both correlators in the $i\ib\to K^{(aa)}\to I^{(ab)}J^{(ab)}$ channel with conformal block $\mathcal{F}_{i\ib IJ}(P_K;m)$, i.e.~expand $\phi_i$ in terms of boundary operators $\phi_{k}^{(aa)}$ and then use boundary-boundary-boundary OPE:
\begin{align}\label{fig:1bulk2bdyope}
\vcenter{\hbox{\includegraphics[height=3cm]{figs/1bulk2bdy.pdf}}} 
=\sum_K~
\vcenter{\hbox{\includegraphics[height=3cm]{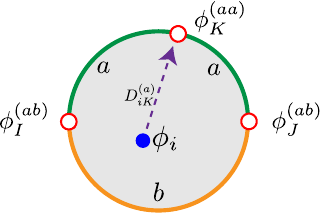}}} ~.
\end{align}
In this channel, taking the average gives
\begin{align}
\overline{G_{iIJ}(m_1) G_{iIJ}^{\prime}(m_2)}
=&\sum_{K,L} \overline{D^{(a)}_{iK}B_{KIJ}^{(aab)}D^{(a)}_{iL}B_{LIJ}^{(aab)}}\mathcal{F}_{i\ib IJ}(P_K;m_1)\mathcal{F}_{i\ib IJ}(P_L;m_2)
\nn
=&\sum_K\overline{\left(D_{iK}^{(a)}\right)^2}\,\overline{\left(B_{KIJ}^{(aab)}\right)^2}\mathcal{F}_{i\ib IJ}(P_K;m_1)\mathcal{F}_{i\ib IJ}(P_K;m_2)
\nn
\approx &\int_0^\infty \d P_K\,\rho_0(P_K)  C_0(i\ib K) C_0(IJK) \mathcal{F}_{i\ib IJ}(P_K;m_1)\mathcal{F}_{i\ib IJ}(P_K;m_2)
\nn
&+ g_a g_b\,\left(D_{i\id}^{(a)}\right)^2\,  \mathcal{F}_{i\ib IJ}(P_\id;m_1)\mathcal{F}_{i\ib IJ}(P_\id;m_2),
\end{align}
where we have used the Gaussian statistics prescribed in Section~\ref{ssec:ensemble} for both $D$ and $B$. In arriving at the final line, we approximated the sum by the integral over the continuum of black hole states plus the identity contribution. Contributions from the sub-threshold non-vacuum contributions are subleading. Notice now that the integral is equal to the Liouville four-point function with complex moduli $(m_1,m_2)$ expanded in the OPE channel $i\ib\to K\to IJ$, which we write as
\begin{align}\label{eq:2copyeg1result}
    \overline{G_{iIJ}(m_1) G_{iIJ}^{\prime}(m_2)} 
    &\approx G_{i\ib IJ}^{\rm L}(m_1,m_2)
    \nn
    &+\left(\sqrt{g_a g_b}D_{i\id}^{(a)}  \mathcal{F}_{i\ib IJ}(P_\id;m_1)\right)\left(\sqrt{g_a g_b}D_{i\id}^{(a)}\mathcal{F}_{i\ib IJ}(P_\id;m_2)\right).
\end{align}
The second term factorizes, but the first does not. It is interesting to point out that the four-point function has its holomorphic cross ratio given by $m_1$ and the anti-holomorphic cross ratio given by $m_2$, but $m_1$ and $m_2$ come from two different copies. In the case of CFTs without borders, the feature is related to the fact that the partition function of the bulk manifold can be calculated by two copies of Virasoro TQFT \cite{CEZ23}. For CFTs with borders, the bulk statement is similar but different \cite{Jafferis:2025yxt}.

\section{3D gravity model}
\label{sec:3dgrav}

In this section, we construct a gravity theory that we expect to reproduce the BCFT ensemble. The ingredients in 3D gravity that we will need involve end-of-the-world (EOW) branes, conical defects, kinks, and punctures. Write the total action as
\begin{align}\label{eq:action_tot}
    I =
    I_{\rm bulk} + I_{\rm conical} + I_{\rm brane} +I_{\rm kink}+I_{\rm puncture}+I_{\rm asymp} .
\end{align}
We will first introduce each term and explain the corresponding object or defect in Section~\ref{ssec:defs}. We then discuss the variational principle in Section~\ref{ssec:var}. An on-shell equivalent but more practical expression will then be presented in Section~\ref{ssec:alt}.

\subsection{Introducing all the defects}\label{ssec:defs}

\begin{figure}
    \centering
    \includegraphics[width=0.8\linewidth]{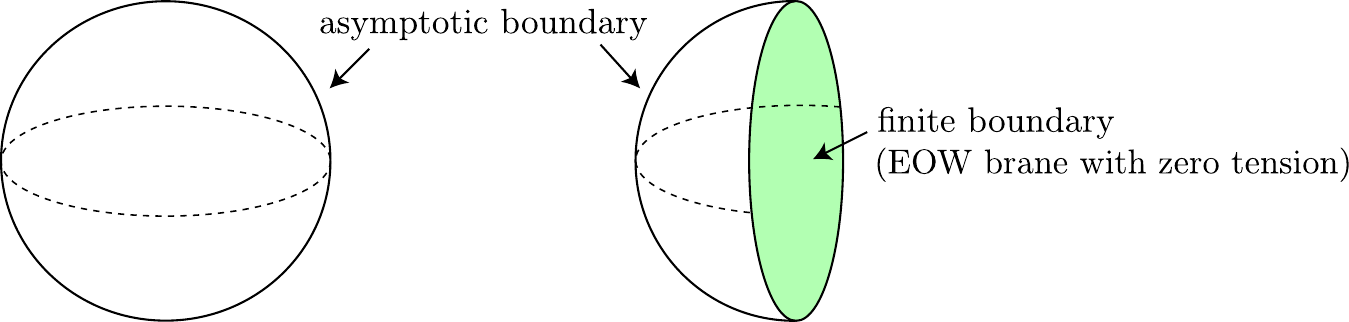}
    \caption{$\mathbb{Z}_2$ quotient of Euclidean AdS$_3$ produces a manifold with an asymptotic disk boundary as well as a finite disk boundary that is a zero-tension EOW brane.}
    \label{fig:Z2quotient}
\end{figure}

The bulk action is the usual Einstein-Hilbert term with a negative cosmological constant,
\begin{align}
    I_{\rm bulk}=-\frac{1}{16\pi G_N} \int_{\mathcal{M}} \sqrt{g} \,(R +2),
\end{align}
where $G_N$ is Newton's constant, and we have set the cosmological constant to be
\begin{align}
    \Lambda = -1,
\end{align}
without loss of generality. Under this convention, the AdS radius is one. Here, $\mathcal{M}$ is a 3D manifold whose boundaries can be purely asymptotic or a combination of asymptotic and finite. For example, Euclidean AdS has one asymptotic boundary, but if we take a $\mathbb{Z}_2$ quotient, we obtain a manifold whose boundary is the union of an asymptotic half-sphere (disk) and a finite disk. See Figure~\ref{fig:Z2quotient}. By ``finite", we mean that its intrinsic geometry has a finite metric; an asymptotic boundary has a finite metric only after an infinite conformal rescaling. 

Next, let us review the action that accounts for conical defects to model bulk scalar primaries with $0<h<c/24$, which is a well-established result in AdS$_3$/CFT$_2$ (see e.g.~\cite{FKW14}). It is given by
\begin{align}\label{eq:act_cone}
    I_{\rm conical}= \frac{1}{2 G_N}\sum_i\int_{{\mathcal{C}_i}} \sqrt{\gamma}\,\eta_i,
\end{align}
where $\gamma_{ab}$ is the induced metric, and $\eta_i$ is related to the corresponding bulk scalar operator dimension in the dual BCFT for each conical defect $\mathcal{C}_i$ via \eqref{eq:etadim}. For example, in Euclidean AdS, one can have a conical defect along an axis of symmetry:
\begin{align}\label{fig:conicaldefect}
   \vcenter{\hbox{\includegraphics[height=3cm]{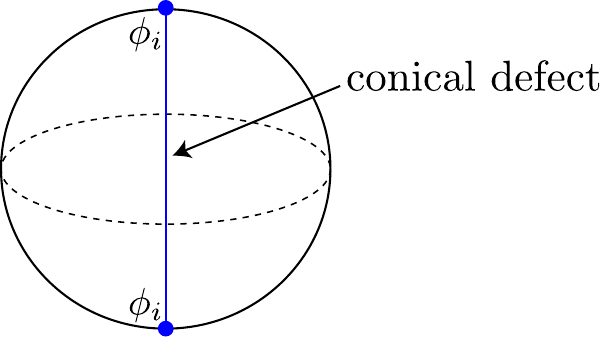}}}  ~.
\end{align}

When we compute observables in the BCFT ensemble such as the example in Section~\ref{ssec:obs}, we will require that these conical defects extend between bulk operator insertions (or end on EOW branes, as we explain later). More precisely, for each conical defect with label $i$, we enforce that it extends from bulk operator $\phi_i$ and ends at another bulk operator $\phi_i$, in either one or two different copies of BCFT. 
Geometrically, since BCFTs live at the asymptotic boundaries of $\CM$, conical defects extend from infinity and are infinitely long. Incidentally, the conical defects model heavy particles in AdS${}_3$, and their trajectories can be obtained from minimizing their actions. 

On shell, the total angle around the defect is $2 \pi(1-2 \eta_i)$, so $\eta_i$ is also proportional to the deficit angle of the on-shell geometry.

In our situation, where the asymptotic boundaries are themselves bordered 2D surfaces, we need some dynamical codimension-one objects in the bulk to account for the extra degrees of freedom from the borders. A natural object to consider is the EOW brane, proposed in \cite{Takayanagi11,FTT11} to model the holographic dual of a single BCFT. The action for the EOW brane $Q$ is given by
\begin{align}\label{eq:action_EOW1}
    I_{\rm brane}=- \frac{1}{8\pi G_N} \int_{Q} \sqrt{h}\, (K-T),
\end{align}
where $h_{ab}$ is the induced metric, $K_{ab}$ is the extrinsic curvature (defined with the normal vector pointing outward), $K$ is the trace of $K_{ab}$, and $T\in(-1,1)$ is a parameter known as the brane tension (see e.g. \cite{KR20}).  This restriction on the range of the brane tension is such that they admit negative-curvature intrinsic geometry, which is a requirement for it to connect to the asymptotic boundaries of $\CM$.\footnote{See \cite{CMW21} for a tighter bound from bootstrap.} We will call EOW branes with tensions within this range ``AdS EOW branes'' because of their locally AdS${}_2$ nature. For example, one can have a 3D manifold whose boundary consists of a disk asymptotic boundary and a disk EOW brane:
\begin{align}\label{fig:eowbrane}
   \vcenter{\hbox{\includegraphics[height=3cm]{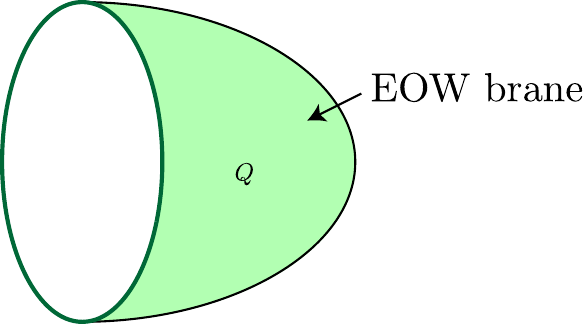}}}  ~.
\end{align}
Like conical defects, the EOW branes can also interpolate between different asymptotic boundaries. The difference is that the branes are 2D objects, and the boundaries of the branes are identified with BCFT borders, i.e., anchored to the BCFT borders. Since BCFT borders have boundary conditions labels, a brane ending on a border with label $a$ should also carry a label $a$, and correspondingly a tension $T_a$. As a rule of our model, we require all the EOW branes to end on the asymptotic boundaries, i.e., compact EOW branes such as spherical branes are not allowed.
Lifting this restriction would result in certain ``one-boundary wormholes", which will be discussed in Section~\ref{ssec:firstmom}. 

So far, this is just the original model used in the AdS/BCFT framework \cite{Takayanagi11,FTT11}, but we will also incorporate branes composed of more than one smooth segment to model the boundary operators, including both boundary-condition-changing ones and boundary-condition-preserving ones. This kind of brane has been previously studied in \cite{MM22,BKSS22} (see also \cite{GLMW21}). On such a brane, $T$ is only piecewise constant, i.e.,~each smooth segment potentially has a different $T$. It reduces to the original EOW brane action when the brane is everywhere smooth and the tension is everywhere constant. We will soon improve the notation to emphasize this point (see \eqref{eq:action_EOW2} below).

Moving on, we introduce \emph{kinks}.\footnote{Below-the-threshold boundary-condition-preserving operators can also be modeled by conical defects \cite{KW22}. We will not adopt this approach because the kink description allows us to treat boundary-condition-changing and boundary-condition-preserving operators on equal footing.} They play a similar role to that of conical defects: A conical defect extends between two CFT bulk operator insertions, while a kink extends between two BCFT boundary operator insertions. Both conical defects and kinks are codimension-two objects in 3D gravity, but the difference is that the kinks must stay entirely on the brane. In other words, from the EOW brane perspective, the kinks are codimension-one defects. As an example, one can have a kink going between two boundary operators on the disk like this:
\begin{align}\label{fig:kink}
   \vcenter{\hbox{\includegraphics[height=3cm]{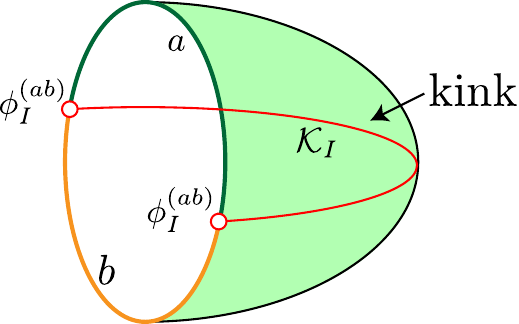}}}.
\end{align}
Their action is given by
\begin{equation}\label{eq:kink}
    I_{\rm kink}=\frac{1}{8 \pi G_N} \sum_I\int_{\mathcal{K}_I}\sqrt{\gamma}\,(\pi-\theta_I),
\end{equation}
where $\theta_I$ is the angle between the branes on the two sides of the kink when measured from within $\mathcal{M}$, and we have again used the notation $\gamma_{ab}$ for the induced metric (because it has the same codimension as the conical defect).\footnote{As we explain in Section~\ref{ssec:alt}, this equals minus the Hayward term \cite{Hayward93}.} 

Notice that we have used the label $I$ for kinks. In Section~\ref{sec:BCFTensemble}, we chose a convention that lower-case letters $i,j,\dots$ labels bulk operators and upper-case letters $I,J,\dots$ label boundary operators. It is no coincidence that we are using lower-case letters for the conical defects and upper-case letters for kinks. As already mentioned, a conical defect labeled by $i$ must connect two bulk operator insertions $\phi_i$ and $\phi_i$ at the asymptotic boundaries. We now require that a kink labeled by $I$ must connect two boundary operator insertions $\phi_I^{(ab)}$ and $\phi_I^{(ab)}$ at the borders of the asymptotic boundaries. Furthermore, since each boundary operator $I$ lives in the Hilbert space {$\mathcal{H}^{(ab)}_{\text{open}}$}, we also require that the corresponding kink must be at the junction between two branes $Q_a$ and $Q_b$, with $Q_a$ extending between borders with the same label $a$, and $Q_b$ extending between borders with the same label $b$, where $a$ and $b$ are generally different, reflected in the fact that the corresponding tensions $T_a$ and $T_b$ are generally different. 

Next, we introduce \emph{punctures}. A puncture is a point-like defect where a conical defect ends on a brane: 
\begin{align}\label{fig:punc}
   \vcenter{\hbox{\includegraphics[height=3cm]{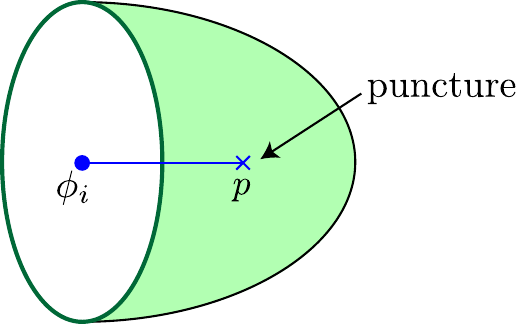}}}  ~.
\end{align}
We will not specify the puncture action here because we find it less distracting to present its action in the alternative form in Section~\ref{ssec:alt}. Among other things, the presence of the puncture ensures $\overline{D_{i\id}^{(a)}}\ne0$, discussed further in Section~\ref{ssec:firstmom}.

Finally, as usual, we have both the Gibbons-Hawking-York term and the counterterm at the asymptotic boundary $\mathcal{A}$ (which can have more than one connected component):
\begin{equation}
    I_{\rm asymp}=- \frac{1}{8\pi G_N} \int_{\mathcal{A}} \sqrt{h}
    \,
    (K-1),
\end{equation}
where we have again used the notation $h_{ab}$ for the induced metric (because it has the same codimension as the brane) and picked the flat conformal frame so that the counterterm that depends on the induced curvature scalar of the cut-off surface is zero.

An example incorporating all the ingredients introduced so far can be found in \eqref{fig:mixedworm}, along with simpler examples in the same section.

\subsection{Variational principle}\label{ssec:var}
Apart from the action, we also need boundary conditions to ensure a good variational principle so that we can determine the solutions. We impose the standard Dirichlet boundary condition at the asymptotic boundary $\mathcal{A}$. At an EOW brane with label $a$, we impose Neumann boundary condition
\begin{equation}\label{eq:braneEOM}
     K_{\mu \nu} = (K-T_a)h_{\mu\nu},
\end{equation}
which we will sometimes refer to as the EOW brane equation of motion. At a conical defect with label $i$, we impose the condition that 
\begin{align}
    \theta_i = 2\pi(1-2\eta_i),
\end{align}
where $\theta_i$ is the conical angle and $\eta_i$ is related to the conformal dimension via \eqref{eq:etadim}. At a kink with label $I$, we fix the kink angle to be $\Theta_I$. The relation between the parameter $\Theta_I$ and the conformal dimension of the corresponding boundary operator $\phi_I^{(ab)}$ is generally complicated, as the kink angle is a function of $g_a$, $g_b$, and $h_I$, i.e., $\Theta_I=\Theta_{I}(h_I,g_a,g_b)$. It simplifies when the kink is in between two tensionless branes, in which case
\begin{align}\label{eq:kinkweight}
    h_I = \frac{c}{24\pi^2}(\pi+{\Theta_{0I}})(\pi-{\Theta_{0I}}),
\end{align}
where $\Theta_{0I}(h_I)\equiv \Theta_I(h_I,0,0)$. We will explain how to determine the relation between $h_I$ and $\Theta_I$ for general $g_a$ and $g_b$ in Section~\ref{sec:gen} and why we only need $\Theta_{0I}(h_I)$ from \eqref{eq:kinkweight} to determine the solution. With these boundary conditions, $I$ has a good variational principle. Incidentally, even though the conformal weights corresponding to the kinks do not appear explicitly in $I$, they enter as boundary conditions. 

\subsection{Alternative action}\label{ssec:alt}
We have explained all terms in the total action and established a good variational principle. However, before moving on to the next section, we find it useful to present a different but on-shell equivalent action.

To explain it, let us first review the following observation: At the location of a conical defect, the Ricci scalar $R$ diverges, but only distributionally \cite{UHIM89}. The Einstein-Hilbert action therefore picks up a finite contribution (per unit length along the conical defect) from an infinitesimal tubular neighborhood of the defect. In Einstein gravity, this contribution can be shown to exactly cancel the action \eqref{eq:act_cone} on shell (see e.g.~\cite{UHIM89,LM13,Dong16,CCHM22}). Therefore, we can equivalently drop this action from the total action if we change the support of the integral $I_{\rm bulk}$ to exclude the defects. We will call this new bulk action $\widetilde{I}_{\rm bulk}$. The variational principle for this alternative formulation is derived in the Appendix A of \cite{DM19}.

Similarly, the integrand of the brane action  \eqref{eq:action_EOW1} is distributional at kinks where $K$ diverges. In the literature, this is often singled out from the smooth sections of the brane and is known as the Hayward term \cite{Hayward93}:
\begin{equation}\label{eq:Hay}
    I_{\rm Hayward}=-\frac{1}{8 \pi G_N} \sum_I\int_{\mathcal{K}_I}\sqrt{\gamma}\,(\pi-\theta_I).
\end{equation} 
This contribution cancels with \eqref{eq:kink}. Together, it is equivalent to another formulation of the kink action presented in \cite{MM22}: 
\begin{equation}\label{eq:actionkink}
    \widetilde{I}_{\rm kink}=-\frac{1}{8 \pi G_N} \sum_I\int_{\mathcal{K}_I}\sqrt{\gamma}\,(\Theta_I -\theta ),
\end{equation}
where the equations of motion sets $\theta=\Theta_{I}$ for each kink $\mathcal{K}_I$, making it vanish on shell.\footnote{In \cite{MM22}, $\Theta_{I}$ does not depend on the brane tensions, so our model is different.}

To understand why the kink action  \eqref{eq:kink} should cancel with the Hayward term, recall that, with the Hayward term alone, the action has a good variational principle if the intrinsic metric at the kink is fixed. However, to satisfy our variational principle, we need the addition of the kink action, which acts like a Legendre transform, to turn the boundary condition from fixing the intrinsic metric to fixing the kink angle.

Given that both $I_{\rm conical}$ and $I_{\rm kink}$ are zero on shell, we can write an action that is equivalent to \eqref{eq:action_tot} on-shell:
\begin{align}\label{eq:action_tot_prime}
    \widetilde{I} =
    \widetilde{I}_{\rm bulk} + \widetilde{I}_{\rm brane}+\widetilde{I}_{\rm puncture}+I_{\rm asymp} ,
\end{align}
where
\begin{align}
    \widetilde{I}_{\rm bulk}=-\frac{1}{16\pi G_N} \int_{\mathcal{M}\backslash\mathcal{C}} \sqrt{g} \,(R - 2\Lambda),
\end{align}
where $\mathcal{C}$ is the union of all conical defects $\mathcal{C}_i$, and
\begin{align}\label{eq:action_EOW2}
    \widetilde{I}_{\rm brane}=- \frac{1}{8\pi G_N} \sum_a\int_{Q_a} \sqrt{h}\, (K-T_a),
\end{align}
where the integrals in the action are only over the smooth segments $Q_a$, which in particular do not include the kinks where $K$ diverges, the label $a$ distinguishes between different segments, and $T_a$ is now a constant parameter for each $a$. We will refer to each $Q_a$ as one brane. In particular, each brane is now smooth, as the kinks are no longer considered to be part of the brane. Punctures are also excluded from the integrand of $\widetilde{I}_{\rm brane}$, and the removal of each puncture introduces a (infinitesimal) circular boundary to the brane. We will come back to this point later in Sections~\ref{ssec:wedgeaction} and \ref{ssec:conj} when we discuss the topological action from the EOW brane.

At this point, we are going to introduce the puncture action:
\begin{align}
    \widetilde{I}_{\rm puncture}=\frac{1}{G_N}\sum_{p} d_p(h_i),
\end{align}
where $d_p$ is some non-zero number we assign to each puncture $p$. The number depends on the weight $h_i$ of the conical defect ending on it. For the purpose of this paper, we will not discuss what these numbers are exactly, although it would be an interesting question to determine whether they are constrained by BCFT bootstrap. 

The asymptotic boundary term is regular at the locations of the conical defects and kinks where they meet the cut-off surface (at any finite cut-off radius), so it does not matter whether they are excluded from the integral. In other words, adding a tilde to $I_{\rm asymp}$ makes no difference.

In the rest of the paper, we will only use $\widetilde{I}$ to evaluate actions because they are much simpler in practice, but since they are equivalent on shell, we will not distinguish them when abstractly referring.

\section{Holographic \texorpdfstring{$g$}{g}-function}
\label{sec:g}

Let us now discuss the $g$-function in more detail, as it plays a key role in later sections. Recall that the $g$-function is given by the disk partition function. To be more specific, the partition function for a disk with boundary condition $a$ is given by $g_a$. However, for notational simplicity, we will often suppress the $a$ subscript when there is only one boundary condition in the context. 

\subsection{\texorpdfstring{$g$}{g} and \texorpdfstring{$T$}{T}}
In AdS/BCFT, to compute the $g$-function in the bulk, one simply needs to find the bulk saddle dual to the empty disk (disk with neither bulk nor boundary operator insertions) on the boundary. Since the bulk saddle is always locally AdS, it can be obtained by removing part of the Euclidean AdS$_3$ geometry:
\begin{align}\label{fig:disk}
        \vcenter{\hbox{\includegraphics[height=3cm]{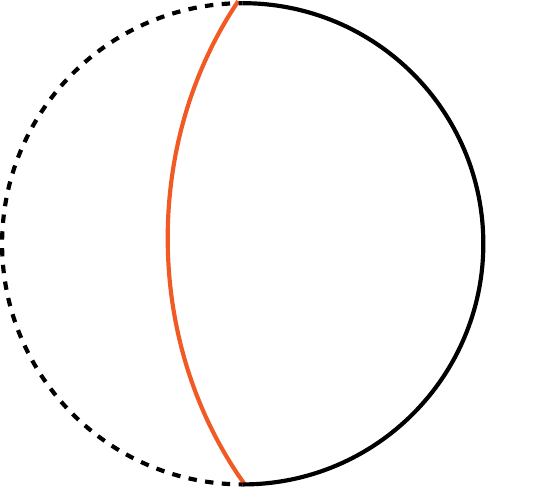}}} 
\end{align}
where the solid black line represents the disk where the BCFT lives, and the EOW brane (orange) cuts off the spacetime so the spacetime to its left no longer exists. The location of the brane is determined by the tension $T$ through the equation of motion
\eqref{eq:braneEOM}, so the total action (both the bulk action and brane action) changes as a function of $T$. Relating the action to the boundary answer $g$ via $g=Z_{\rm disk}=e^{-I}$ gives \cite{Takayanagi11,FTT11}
\begin{align}\label{eq:gandT}
    \log g = \frac{1}{4G_N}{\rm arctanh} \,T= \frac{c}{6}{\rm arctanh} \, T. 
\end{align}
More generally, for different boundary conditions $a$, each $g$-function $g_a$ is related to the corresponding function $T_a$ via the same equation. 

There is an important difference between AdS/BCFT and ``AdS/BCFTs" as far as the $g$-function is concerned. As long as there is one tunable parameter in the EOW brane action (or boundary condition), we can establish a relation between that parameter and the $g$-function by insisting that the on-shell partition functions of the disk agree. The relation between $g$ and $T$ depends on the choice of the action, but there is no non-trivial consistency condition on this relation. 

However, when we consider ensemble observables, we get non-trivial conditions on the EOW brane action. For example, consider the two-copy observable \eqref{eq:2copyeg1}. From the result \eqref{eq:2copyeg1}, we can see that it has no dependence on the $g$-functions $g_a$ and $g_b$. Since the relation between $g$ and $T$ is already fixed in \eqref{eq:gandT}, we no longer have any parameters in the 3D model to tune. Requiring that the corresponding bulk saddle (presented later in \eqref{fig:1bk2bdworm}) be independent of $g$-functions is then a constraint on the form of the EOW action. Similarly, we can consider two-copy observables of other correlators and even many-copy observables (which generally depend on the $g$-functions in a non-trivial but simple way), so that we get infinitely many consistency tests that the choice of the EOW brane action must pass. 

Checking that the action in Section~\ref{sec:3dgrav} indeed passes all such tests is an important result that we establish fully when there are no kinks (Section~\ref{sec:tension}) and partially when there are kinks (Section~\ref{sec:gen}).

\subsection{A no-ensemble theorem for \texorpdfstring{$g$}{g}}
\label{ssec:nogo-g}
According to some evidence \cite{SW22,CCHM22}, the low-energy spectrum ($h\sim c^0$) is not prone to averaging (in holography), so quantities such as $\overline{D_{iI}D_{iI}}$ and $\overline{B_{IJK}B_{IJK}}$ are expected to factorize when all the operator labels are set to the identity $\mathbbm{1}$, i.e., $\overline{g_a^2}=g_a^2$.\footnote{Our results meet this expectation, but the general principle beyond 3D may need refinement \cite{ASS23,AR24}.} In other words, the $g$-functions are not expected to exhibit averaging and should be treated as constants from the BCFT ensemble perspective. 

In gravity, this translates to the statement that there should be no wormholes connecting two empty disks. With the ensemble interpretation, such wormholes contribute to the variance of $g_a$. If the variance of a probability distribution is zero, the variable must be a constant. 

As a consequence of Witten-Yau theorem \cite{WY99,CG99}, wormholes with metric of the form 
\begin{equation}\label{eq:MMmetric}
    \d s^2 = \d\rho^2 + \cosh^2\rho \,\d\Sigma^2
\end{equation}
only admit solutions if $\Sigma$ admits a negative-curvature metric. These solution are also known as Maldacena-Maoz wormholes \cite{MM04}. If $\Sigma$ is topologically a sphere, such a solution is therefore forbidden. Topological disks, on the other hand, do admit hyperbolic metrics, so the Witten-Yau theorem does not forbid the existence of wormholes connecting two empty disks. Nonetheless, we will show that there are no wormholes connecting two empty disks 
while preserving the $U(1)$ symmetry inherited from the disk, both by providing an argument and by an explicit calculation.

\subsubsection*{Argument: traversable wormhole}
As an explicit calculation will be presented next, the argument we present here does not add to the results, but we find it useful for narrative reasons and for the benefit of providing a comparison to the construction of \cite{CHM24} regarding Liouville line defects. 

Suppose that there is an on-shell wormhole configuration $\CM$ of the action \eqref{eq:action_tot} connecting two empty disks, while preserving the $U(1)$ symmetry inherited from the disks. Since a disk is conformally related to a half plane, we can perform a coordinate transformation on the wormhole such that it now connects two half planes, say $x>0$ in the $(x,\tau)$ plane, and the wormhole will have a translational symmetry along the $\tau$ direction. Let $x=0$ denote the location of the EOW brane everywhere in the bulk. We can then glue this wormhole to a copy of itself along $x=0$, so that this doubled manifold which we will denote as $\mathcal{M}'$ has the topology of $\mathbb{R}^2$ times an interval. At $x=0$, however, the metric is continuous but not differentiable (unless the brane tension is zero). This codimension-one object is often known as a \emph{thin brane} \cite{Israel66}, which, unlike the EOW brane, has spacetime on both sides.

The total Euclidean action for $\mathcal{M}'$ is now the sum of the original actions,
    \begin{align}
        I_{\rm bulk}+I_{\rm thin} = -\frac{1}{16\pi G_N} \int_{\mathcal{M}'} \sqrt{g}\, (R + 2) - \frac{1}{8\pi G_N} \int_Q \sqrt{h}\, (K_{\rm L} - K_{\rm R} - 2T),
    \end{align}
where $K_{\rm L}$ and $K_{\rm R}$ are the extrinsic curvature of the thin brane $Q$ as measured from its two sides, which we call left and right, and they are both defined with the normal that points from left to right. 

From the variational principle, 
\begin{equation}
    \delta I_{\rm thin} = - \frac{1}{8\pi G}\int_Q \sqrt{h} \left(\frac{1}{2}(h_{\mu \nu}[K]-[K_{\mu \nu}])-T h_{\mu \nu}\right) \delta h^{\mu \nu},
\end{equation}
where $[K_{\mu \nu}]=K_{\text{L},\mu \nu}-K_{\text{R},\mu \nu}$ is the jump of the extrinsic curvature at $Q$, we obtain the Israel junction condition
\begin{equation}
    h_{\mu \nu}[K]-[K_{\mu \nu}]=2Th_{\mu \nu}.
\end{equation}
The full equation of motion is
\begin{equation}
    E_{\mu \nu}\equiv -\frac{16\pi G_N}{\sqrt{g}}\frac{\delta I_{\rm bulk}}{\delta g^{\mu \nu}}=\frac{16\pi G_N}{\sqrt{g}}\frac{\delta I_{\rm thin}}{\delta g^{\mu \nu}}\equiv 8\pi G_N T_{\mu \nu},
\end{equation}
with the stress tensor given by
\begin{equation}
    T_{\mu \nu}=\frac{2}{\sqrt{g}}\frac{\delta I_{\rm thin}}{\delta g^{\mu \nu}} = \frac{1}{8\pi G_N}\left(-[K]h_{\mu \nu}+2T h_{\mu \nu}\right)\delta(x),
\end{equation}
where $h_{\mu \nu}=g_{\mu \nu}-n_\mu n_\nu$ is the metric projected onto the thin brane, and $x$ is the proper distance along geodesics orthogonal to the thin brane $n_\mu=\partial_\mu x$, and we pick $x$ such that the thin brane is located at $x=0$.  On-shell, i.e., after using the Israel junction condition, it simplifies to
\begin{equation}
    T_{\mu \nu}=-\frac{1}{4\pi G_N}T h_{\mu \nu}\,\delta(x).
\end{equation}

The advantage of relating an EOW configuration $\CM$ to a thin brane configuration $\CM'$ is that $\CM'$ is a more familiar manifold, where the thin brane can be thought of as some matter field densely packed at a dimension-one surface. Continuing with the argument, we now perform an analytic continuation $\tau\to \i t$, where $\tau$ is the Euclidean coordinate orthogonal to $x$ (parallel to the brane). Since the original Euclidean wormhole has translation symmetry in $\tau$, this analytic continuation will turn $\CM'$ into a spacetime with both time-translation symmetry and a $\mathbb{Z}_2$ symmetry ($x\to-x$). It therefore has the following light ray going from one asymptotic boundary to the other:
\begin{align}\label{fig:traverse}
        \vcenter{\hbox{\includegraphics[height=4cm]{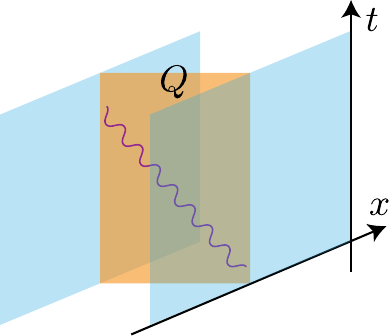}}} 
\end{align}
where two asymptotic boundaries are represented as blue planes, the thin brane $Q$ (orange) is a timelike hypersurface connecting two boundaries at $x=0$, and the purple wiggly line represents a light ray on the brane. Assuming that any infinitesimal deformation to this light ray also remains one that connects the two boundaries, this makes the spacetime an eternally traversable wormhole. It is unclear whether this assumption holds, making it a potential loophole in the argument.

For $T\ge0$, the spacetime satisfies Null Energy Condition (NEC) everywhere, because for any null vector $V^\mu$,
\begin{equation}
    T_{\mu \nu}V^\mu V^\nu=-\frac{T}{4\pi G} (g_{\mu \nu}-n_\mu n_\nu) V^\mu V^\nu \delta(x)=\frac{T}{4\pi G} (n_\mu V^\mu)^2\delta(x)\ge0,
\end{equation}
where $n^\mu$ is the spacelike normal vector of the thin brane that is normalized to $n^\mu n_\mu =1$. Since we now have a traversable wormhole satisfying NEC, which is forbidden by the topological censorship \cite{FSW93}, we have reached a contradiction. 

Note that the argument above does not immediately rule out wormholes in the $T<0$ case, since there NEC is violated. Plus, we have a potential loophole in the argument. Therefore, we would like to present an explicit calculation that applies to both signs of $T$.

Before proceeding, let us make a comment on a similar object called thin shells, which are like thin branes but with Euclidean perfect fluids localized on them. Consider for example the Euclidean wormholes in \cite{CM21,BLMS22,CHM24} supported by thin shells, and perform analytic continuation in the $U(1)$ symmetric direction. Along the shells, light rays can traverse from one asymptotic boundary to another, which is possible because these shells violate NEC. A similar situation has been noticed in \cite{KKSTTW23}, where a scalar field localized on the EOW brane \cite{KSSTW23} was considered. Since these Euclidean wormholes and therefore their analytically continued Lorentzian wormholes exist, the assumption that small deformations of those traversable light rays remain traversable must be invalid. It would be interesting to verify this.\footnote{We thank Don Marolf for discussion on this point.}

\subsubsection*{Calculation: Maldacena-Maoz wormhole}
We now perform an explicit calculation to show that there are no wormholes connecting two empty disks if we assume that a solution takes the form of a Maldacena-Maoz wormhole \eqref{eq:MMmetric} with a hyperbolic metric on $\Sigma$: 
\begin{equation}
    \d s^2 = \d\rho^2 + \cosh^2\rho \left( 4\frac{\d r^2+r^2\d\theta^2}{(1-r^2)^2}\right).
\end{equation}
where the metric on each constant $\rho$ slice is a hyperbolic disk, with $r=1$ being its conformal boundary in polar coordinates $(r,\theta)$; $\rho=\pm\infty$ are the two asymptotic boundaries of the wormhole. Let the location of the EOW brane be given by a to-be-determined positive function $r=f(\rho)$, then boundary conditions require that $r \to r_0$ as $\rho \to \pm \infty$, for some $r_0<1$. Assuming $\mathbb{Z}_2$ symmetry $\rho\to-\rho$, we also require $f'(0)=0$.

For this choice of coordinates, consistency of the two components of the brane equation of motion \eqref{eq:braneEOM}
requires either $f(\rho)=1$, which is a trivial solution, or
\begin{equation}
    f''(\rho)+ f'(\rho)^2\frac{(1+3f(\rho)^2)}{f(\rho)(1-f(\rho)^2)}+ f'(\rho)\tanh\rho+\frac{(1-f(\rho)^4)}{4f(\rho)\cosh^2\rho}= 0.
\end{equation}
The non-trivial solution to the above equation is
\begin{equation}
    f(\rho)=\sqrt{\frac{4c_1 c_2+\cosh\rho +4c_1\sinh\rho}{4c_1 c_2-\cosh\rho +4c_1\sinh\rho}}.
\end{equation}
for some undetermined constants $c_1,c_2$. However, there does not exist a choice of $c_1,c_2$ such that $f'(0)=0$ and the boundary conditions for $f(\r)$ can be satisfied at the same time.

In summary, the brane equation of motion forbids the existence of two-boundary empty wormholes, at least if we assume $U(1)$ symmetry. From now on, we will treat $g$-functions as constants in the ensemble interpretation of BCFT data. 

As a bonus, it immediately follows from this calculation that there does not exist a solution connecting two punctured disks (disks with a bulk operator insertion), assuming $U(1)$ symmetry in $\theta$, as the calculation goes through if we change the periodicity of $\theta$ from $2\pi$ to some conical angle smaller than $2\pi$. Using the ensemble language, $\overline{D^{(a)}_{i\id}D^{(a)}_{i\id}}|_{\text{conn.}}=0$ for sub-threshold states $i$, consistent with our choice of the ensemble \eqref{eq:Diid2}. (For black hole states, there does exist an on-shell configuration. It can be visualized by doing the half-toroidal surgery described in Section~\ref{eq:cross} on the conical defect $i$ in the would-be wormhole contributing to $\overline{D^{(a)}_{i\id}D^{(a)}_{i\id}}|_{\text{conn.}}=0$ with a sub-threshold $i$. Its action can be found in \cite{Takayanagi11}.)

\section{Wormholes with no tension}
\label{sec:tensionless}

We are now ready to discuss wormhole solutions of our gravity model presented in Section~\ref{sec:3dgrav}. First of all, we want to emphasize that this model is an extension of \cite{CCHM22}: All solutions of 3D gravity, pure or with conical defects, are still solutions of our model. They still compute statistical moments of 
$C_{ijk}$ in our ensemble and more generally CFT correlators of bulk operators on closed Riemann surfaces. One may wonder whether we can have a simpler setup by studying only wormholes connecting bordered Riemann surfaces and still have a consistent duality between these solutions and a simpler ensemble involving less data, say without $C_{ijk}$. However, even an object as simple as the disk partition function with two bulk operator insertions and one boundary operator insertion can be expanded in a channel that involves $C_{ijk}$, so the $C$'s, $B$'s and $D$'s cannot be divided into independent sectors. In fact, in Section~\ref{ssec:mix}, we will discuss wormholes that compute statistical moments involving more than one type of OPE coefficients.

Before discussing the most general solutions of Section~\ref{sec:3dgrav}, it will be very instructional to discuss the special case where all the tension parameters $T_a$ are set to zero. This is what we will focus on in this section. As we will see in later sections, the construction of wormholes with tensionless EOW branes turns out to be a pivotal step even in the generally tensionful case.

\subsection{Quadrupling trick}
\label{ssec:quadr}
To simplify computations in a BCFT, Cardy introduced a method known as the doubling trick to relate the computation of a correlator on a surface with boundary to a correlator on a surface without boundary \cite{Cardy84}. 

Start with a Riemann surface $\Sigma_{g,n}$ with genus $g$ and $n>0$ borders. We can obtain a compact Riemann surface by gluing to it a mirror image of itself along the $n$ boundaries. The resulting manifold is denoted by $\Sigma'_{g',0}$, which now has genus $g'=2g+n-1$ and no borders. For example, a disk is doubled to a sphere, and a genus-one surface with two boundaries is doubled to a genus-three surface. Regarding operator insertions in $\Sigma_{g,n}$, for every bulk operator with weights $(h_i,\hb_i)$, one keeps the chiral half fixed and sends the anti-chiral half to its mirror image, while leaving the boundary operators unchanged. The resulting object is the correlator of many bulk operators in a \emph{chiral} CFT.

We do not want to deal with chiral CFTs because we are working with 3D gravity in the bulk which is non-chiral. We will therefore do a slightly different doubling trick, which is a ``doubled" version of the doubling trick, which we will call the \emph{quadrupling trick}. The doubled Riemann surface is obtained in the same way, but for operators, we make a copy of each bulk operator at its mirror location and turn each boundary operator of dimension $h_I$ to a bulk operator with $h_i=\hb_i=h_I$. An important difference between the quadrupling trick and the original Cardy's version of the doubling trick is that our procedure does not compute the same correlator, while the original version does. This means that we need to halve the result at some point to ensure that we do not overcount. Note that the quadrupling trick does not work for correlators of spinning operators, but it works for us because we only consider scalar operator insertions. 

\subsection{Halving trick}\label{ssec:halve}

Once we have performed the quadrupling trick for each BCFT correlator of interest, we can find wormholes connecting them, which do not involve branes or kinks, just as the ones in \cite{CCHM22}. For each $\CM'$ connecting them, we obtain a contribution
\begin{align}
    \langle Z(\Sigma'_{g'_1,0})\cdots Z(\Sigma'_{g'_m,0})\rangle\supset Z(\CM').
\end{align}

We then look for manifolds $\CM'$ with a $\mathbb{Z}_2$ symmetry that reduces to the $\mathbb{Z}_2$ symmetry on each $\Sigma'_{g',0}$ that takes each point to its mirror image along the gluing surface. In other words, quotienting $\CM'$ by this $\mathbb{Z}_2$ also turns each of its asymptotic boundary $\Sigma'_{g',0}$ back to $\Sigma_{g,n}$. We will refer to the procedure of taking this quotient as the \emph{halving trick} and the resulting manifold as $\CM$. Unlike the doubling trick which is a BCFT technique, the halving trick is a procedure in 3D gravity.

We now claim that the orbifold $\CM$ obtained this way is an on-shell configuration to the action \eqref{eq:action_tot} with all tensions set to zero, as long as we also place EOW branes at the $\mathbb{Z}_2$ fixed points. 

First of all, the bulk equation of motion is obviously satisfied because it is still locally AdS${}_3$ everywhere. Secondly, the conical equation of motion is similarly satisfied for all conical defects away from the EOW branes. Moreover, at every $\mathbb{Z}_2$ fixed points of $\CM'$ away from conical defects, $K_{ab}=0$, so the EOW equation of motion \eqref{eq:braneEOM} is satisfied. Finally, we need to deal with the remaining issue of conical defects that are located at $\mathbb{Z}_2$ fixed points. 

When a conical defect lies exactly on the $\mathbb{Z}_2$-symmetric surface, taking the quotient turns the geometry in its neighborhood into a corner, in the sense that the normal vector at the EOW brane changes discontinuously. As defined in Section~\ref{sec:3dgrav}, we call it a kink. For the geometry near the kink to satisfy the equation of motion, we need to make sure that the angle it has (which is half of the conical angle of the defect in $\CM'$) is compatible with the action \eqref{eq:actionkink} whose parameters are fixed by the corresponding operators at the asymptotic boundaries. This is in fact true by construction, i.e., we have derived \eqref{eq:kinkweight} by requiring this to be true. 

When a conical defect is orthogonal to the $\mathbb{Z}_2$-symmetric surface, we obtain a puncture at the intersection between the conical defect and the $\mathbb{Z}_2$-symmetric surface. The action of the puncture is a number, so there are no additional equations of motion to check.

Finally, we need to ensure we get the correct action. This is where the alternative form of the action \eqref{eq:action_tot_prime} proves particularly convenient. Recall that in this form, there is no contribution from conical defects or kinks, and in the case of zero tension, no contributions from the EOW branes either (because $K=T=0$). The action of $\CM$ is therefore half of that of $\CM'$. It then follows that
\begin{align}
    \langle Z(\Sigma_{g_1,n_1})\cdots Z(\Sigma_{g_m,n_m})\rangle \supset Z(\CM)=\sqrt{Z(\CM')}.
\end{align}

\subsection{Examples}
\label{ssec:mix}

While the tricks we just explained work in general, it is useful to see them in action in various different scenarios. We will first give some examples computing the statistical moments of the fundamental data $C$, $B$, and $D$. We then give an example that computes a more complicated two-copy observable in the ensemble where the two copies can have different moduli. Next, we discuss scenarios where the wormhole computes mixed moments between data $C$, $B$, and $D$. Finally, we will discuss the special scenario of the moments. In all the examples, we will assume that the EOW branes all have zero tension, so that the quadrupling and halving tricks work. Adding tension will be the role of the next two sections.

An important example in 3D gravity with conical defects is the $C^2$ wormhole which computes the second moment of the ensemble. It looks like
\begin{align}\label{fig:C0}
    \vcenter{\hbox{\includegraphics[height=2cm]{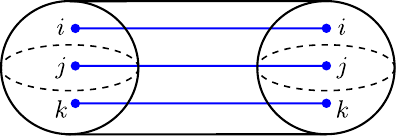}}}
\end{align}
where the boundaries are three-punctured spheres and the blue lines represent conical defects. As explained in \cite{CCHM22}, this corresponds to $|C_0(ijk)|^2$ (up to a conformal transformation). 

We now arrange the location of the conical defects in this wormhole so that they all lie on the equator of the sphere. The wormhole then has a $\mathbb{Z}_2$ symmetry. Quotienting by it gives what we will call the $B^2$ wormhole
\begin{align}\label{fig:B2worm}
    \vcenter{\hbox{\includegraphics[height=2cm]{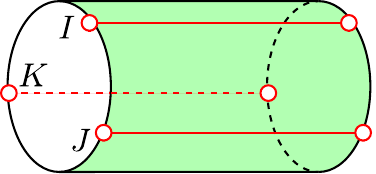}}}
\end{align}
where the boundaries are now disks, the green annular surface is the surface of $\mathbb{Z}_2$ fixed point which we identify as the EOW brane, and the red lines represent kinks. As explained in the previous subsection, the action of this wormhole is half of the $C^2$ wormhole, so its partition function is the square root of $|C_0(ijk)|^2$. As $h_i=\hb_i=h_I$, etc.~(by construction), the result of taking the square root is just $C_0(IJK)$, producing the second moments of $B$ in the BCFT ensemble described in \eqref{eq:B_second}. 

We can alternatively arrange the location of the conical defects in \eqref{fig:C0} in the following way: $i$ at the north pole, $j$ at the south pole, and $k$ on the equator. Furthermore, let us choose $i$ and $j$ to have the same defect angle. Taking the quotient of the $\mathbb{Z}_2$ symmetry that reflects every point around the equator gives another wormhole
\begin{align}\label{fig:D2worm}
    \vcenter{\hbox{\includegraphics[height=2cm]{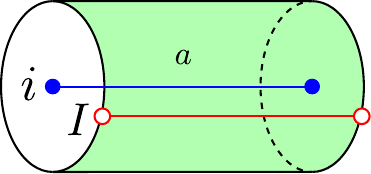}}}
\end{align}
where the boundaries are again disks, the green annular surface is the EOW brane, the red line is a kink, and the blue line remains a conical defect. We call this the $D^2$ wormhole. After identifying $h_k=\hb_k=h_I$, the partition function is given by the square root of $C_0(iiI)C_0(\ib\ib I)$ which is $C_0(iiI)$, in agreement with \eqref{eq:D_second} (recall that conical defects have $h_i=\hb_i$). 

This procedure easily generalizes to higher moments of $B$ and $D$. Taking a multi-boundary wormhole connecting four or more boundaries, each being a three-punctured sphere, quotienting the geometry by a $\mathbb{Z}_2$ symmetry straightforwardly gives $\overline{B\cdots B}$ and $\overline{D\cdots D}$. One way to obtain them is from Virasoro TQFT diagrams with Wilson lines connecting three-punctured spheres (see e.g.~\cite{CEZ24,dBLP24}), though one should be aware that not all TQFT diagrams can be turned into classical multi-boundary wormholes connecting three-punctured spheres by analytically continuing the weights of the Wilson lines below the threshold. Explicit constructions of wormholes connecting $B$'s and $D$'s can be found in \cite{Jafferis:2025yxt}.

Let us move on to a scenario where the boundaries have moduli. Take the two-copy observable \eqref{eq:2copyeg1} whose ensemble-averaged expression was given by \eqref{eq:2copyeg1result}. Let us determine the contributing wormhole. Doing the quadrupling trick, we turn each disk 1-bulk-2-boundary function into a sphere 4-point function of four scalar bulk operators. The wormhole saddle looks like
\begin{align}\label{fig:4ptworm}
    \vcenter{\hbox{\includegraphics[height=2cm]{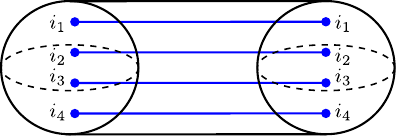}}}
\end{align}
which connects two four-punctured spheres. All the details about this wormhole can be found in \cite{CCHM22}, but we only need the action, which is given by a simple formula in terms of the Liouville four-point function $G^{\rm L}_{i_1i_2i_3i_4}(m_1,m_2)$:
\begin{align}
I(\CM')= -\log \left|G^{\rm L}_{i_1i_2i_3i_4}(m_1,m_2)\right|^2,
\end{align}
where $\phi_{i_1}$ etc.~are all scalar operators. Because of the quadrupling trick, the locations of these four scalar insertions on the sphere have a $\mathbb{Z}_2$ symmetry. Identify $i_1$ with the original operator insertion $i$ and $i_2$ with the mirror image of $i$; the remaining $i_3$ and $i_4$ are bulk operators with weights $h_3=\bar{h}_3=h_I$ and $h_4=\bar{h}_4=h_J$. With this arrangement, we only have two real moduli parameters $m_1=\bar{m}_1$ and $m_2=\bar{m}_2$. 

The halving trick tells us to take a $\mathbb{Z}_2$ quotient of this wormhole along the equator. The resulting manifold $\CM$ then looks like
\begin{align}\label{fig:1bk2bdworm}
    \vcenter{\hbox{\includegraphics[height=2cm]{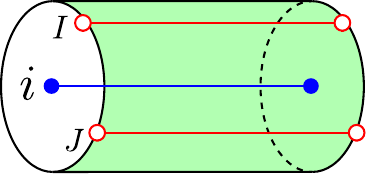}}}
\end{align}
where the asymptotic boundaries are both disks with one bulk and two boundary operators but have different moduli $m_1$ and $m_2$. Since the action is halved by the quotient, 
\begin{align}
    Z(\CM)
    =G^{\rm L}_{i\ib IJ}(m_1,m_2).
\end{align}

It is also possible to obtain wormholes without kinks. As an example, take a Maldacena-Maoz wormhole connecting two tori, each with two bulk insertions. Placing the two bulk operators in a $\mathbb{Z}_2$-symmetric manner and cutting it in half gives
\begin{align}\label{fig:torusworm}
    \vcenter{\hbox{\includegraphics[height=2cm]{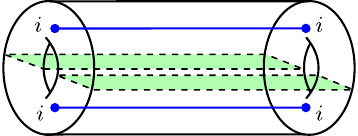}}}
\end{align}
where the two green planes schematically represent a $\mathbb{Z}_2$-symmetric cut, each with topology $S^1\times \mathbb{R}$. The resulting orbifold computes the two-copy observable of annulus one-bulk-point functions. 

A novel feature of the BCFT ensemble is that the different types of data $C$, $B$, and $D$ have joint probability distributions, i.e.,~their moments do not factorize. In the language of gravity, this is manifested as the existence of wormholes like the following (where only boundaries of the wormhole are drawn):
\begin{align}\label{fig:mixedworm}
    \vcenter{\hbox{\includegraphics[height=4cm]{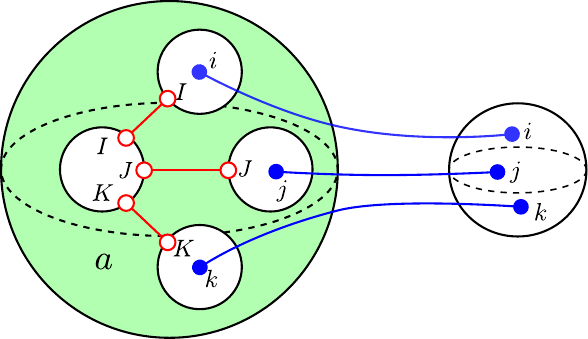}}}
\end{align}
whose boundary has two connected components of spherical topology: one entirely asymptotic representing data $C_{ijk}$, the other composed of a finite piece representing the EOW brane (green) and four asymptotic disks. The disk with three boundary operator insertions represents the data $B_{IJK}^{(aaa)}$, and the disks with one bulk and one boundary insertions represent $D_{iI}^{(a)}$, etc. As in previous diagrams, the blue lines represent conical defects, and the red lines represent kinks. This can be obtained from taking the $\mathbb{Z}_2$ quotient of a wormhole connecting six spheres, with nine conical defects. Working out the action of these wormholes from the classical gravity perspective is generally difficult, but it can be obtained rather simply using Virasoro TQFT \cite{CEZ23}. More examples like this are presented in \cite{Jafferis:2025yxt}. 

Up to this point, all the kinks and conical defects extend to the asymptotic boundaries. The simplest example that does not have this feature is the $\mathbb{Z}_2$ quotient of Euclidean AdS${}_3$ with a conical defect extending between the north and south poles:
\begin{align}\label{fig:sphere2pt}
    \vcenter{\hbox{\includegraphics[height=3cm]{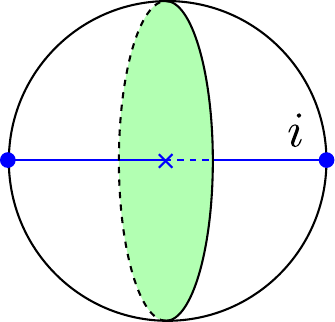}}}\quad\xrightarrow{\mathbb{Z}_2}\quad 
    \vcenter{\hbox{\includegraphics[height=3cm]{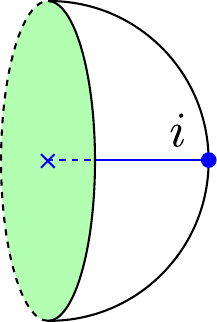}}}
\end{align}
where the manifold on the left computes the sphere two-point function, which is just a normalization, while the manifold on the right computes $D_{i\id}^{(a)}$ (up to a conformal transformation).

Another more interesting example is can be obtained from a wormhole with six conical defects
\begin{align}
    \vcenter{\hbox{\includegraphics[height=2cm]{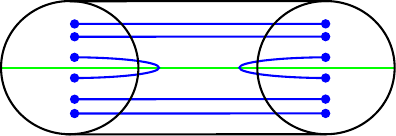}}}
    \quad \xrightarrow{\mathbb{Z}_2}
    \quad \vcenter{\hbox{\includegraphics[height=2cm]{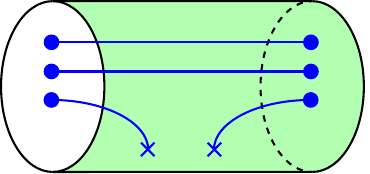}}}
\end{align}
where there are now two punctures on the EOW brane. This contributes to the two-copy observable of three-bulk-point correlator on the disk in an addition to the wormhole with three conical defects all extending between the disks.
\section{Wormholes with no kinks}
\label{sec:tension}
We have explained how to obtain all wormhole solutions with tensionless EOW branes. Before considering the most general case, which will be the topic of the next section, we find it illuminating to first present results on wormholes with no kinks but allowing for arbitrary tension, conical defects, and punctures.

As we will see in Section~\ref{ssec:wedgetrick}, such wormholes can be simply constructed by gluing a wedge-shaped region of spacetime to the wormholes with (possibly punctured) tensionless branes constructed before. We will then compute their actions in Section~\ref{ssec:wedgeaction} and find that the action of the wedge region is topological.

Negative tensions are more subtle than positive tensions, so we will restrict to $T>0$ for most of the section. Nevertheless, negative tensions can be interesting, and we will have a short discussion in Section~\ref{ssec:negtension}.

\subsection{The wedge trick}
\label{ssec:wedgetrick}
Given a solution obtained by $\mathbb{Z}_2$ quotient, we want to determine the solution after we change the tension in the action \eqref{eq:action_EOW1}. After all, the tension parameters $T_a$ are not up to us: They are determined by the $g$-functions $g_a$ corresponding to different boundary conditions labeled by $a$ in the BCFT. 

We can already get some hints from the defining example in Section~\ref{sec:g}. Consider again the geometry in \eqref{fig:disk}. If the brane is tensionless, then the geometry is just half of the Euclidean AdS. Starting with half of the Euclidean AdS, we can obtain the desired solution when the tension is non-zero by gluing a wedge to it:
\begin{align}\label{fig:diskwedge}
    \vcenter{\hbox{\includegraphics[height=2cm]{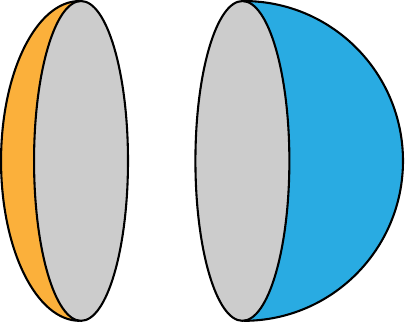}}} 
\end{align}
where the blue disk represents the asymptotic boundary and the orange disk is the positive-tension brane. They are then glued along the grey surfaces.

This procedure works more generally, i.e., we can always obtain the solution for a positive-tension brane by gluing a wedge to the solution with the tensionless brane. It is easy to write down the metric of such a wedge:
\begin{align}\label{eq:coorwedge}
    \d s^2 = \d \rho^2 + \cosh^2\rho\, \d s^2_{\Sigma},
\end{align}
where $\d s^2_{\Sigma}$ is the intrinsic metric of the tensionless EOW brane $\Sigma$. Because $\Sigma$ is the $\mathbb{Z}_2$ fixed points, it has zero extrinsic curvature and is locally AdS${}_2$ with its AdS${}_2$ length equal to the AdS${}_3$ length (which is $L=1$). We define $\rho$ to be increasing as the EOW brane is approached from inside. We choose $\rho=0$ to be where the brane would be located if $T=0$; for $T>0$, the brane is located at some $\rho_*>0$ whose value is to be determined. The surface $\Sigma$ can have an arbitrary genus, an arbitrary number of boundaries, and an arbitrary number of punctures.

We can directly check that this metric is locally AdS${}_3$, so it satisfies Einstein's equation. From the action \eqref{eq:action_EOW1}, we can figure out the variational principle at the brane. Imposing Neumann boundary condition leads to
\begin{equation}
    K_{\mu\nu}-(K-T) h_{\mu\nu}=0.
\end{equation}
The extrinsic curvature at each fixed $\rho$ is given by
\begin{align}
    K_{ij}=\tanh\,\rho\, h_{ij},
\end{align}
with trace
\begin{align}
    K= K_{ij}h^{ij}= 2\,\tanh\,\rho.
\end{align}
At the EOW brane $Q$ where $\rho=\rho_*$, Neumann boundary condition then enforces that $\rho_*$ should be related to the tension via
\begin{align}\label{eq:Tandrhostar}
    T=  \tanh \,\rho_* . 
\end{align}

This shows that the manifold obtained by the adhesion of a wedge is the desired wormhole solution. For example, if we have a wormhole with tensionless EOW branes connecting two disks (supported by some conical defects), obtained by taking the $\mathbb{Z}_2$ quotient of a Maldacena-Maoz wormhole, the wormhole with tension is obtained by gluing
\begin{align}
    \vcenter{\hbox{\includegraphics[height=2cm]{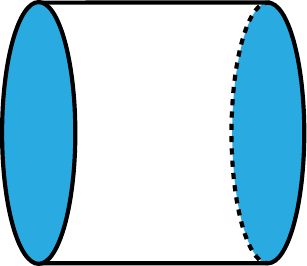}}} \quad +\quad   
    \vcenter{\hbox{\includegraphics[height=0.5cm]{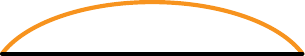}}}\times S^1
\end{align}
where the blue disks are asymptotic boundaries and the orange annulus is the tensionful brane.

\subsection{Tension is topological}
\label{ssec:wedgeaction}
Now that we know how to obtain the general wormhole solutions with tensionful branes from ones with tensionless branes, let us determine how the action changes as a function of the tension. Some simple examples were studied in \cite{AKTW20}, and the general case without kinks or punctures was discussed in \cite{Geng22}. In this section, we revisit the analysis of \cite{Geng22}, now incorporating a treatment of the boundary terms at the asymptotic regions.  We will show that the total action of the wedge region solely depends on the brane tension $T$ and the topology of $\Sigma$, by demonstrating that $I_{\rm asymp}$ exactly combines with $I_{\rm brane}$ and $I_{\rm bulk}$ into a topological expression.

\begin{figure}
    \centering
    \includegraphics[width=0.5\linewidth]{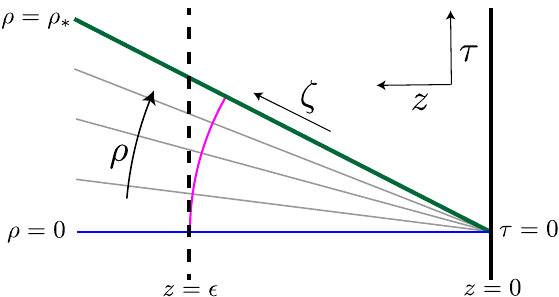}
    \caption{Geometry of the 3D manifold near an asymptotic boundary ($z=0$). The EOW brane $Q$ (green) meets the boundary at an angle determined by the tension. The direction $\theta$ is orthogonal to the plane of the paper. If the tension were zero, the EOW would be located at $\rho=0$, which we label $\Sigma$ (blue horizontal line). The pink arc is where $\zeta=\epsilon$. The full geometry is cut off at $z=\epsilon$ (dashed vertical line). The region between $\Sigma$ and $Q$ is foliated by constant-$\rho$ slices, each locally AdS${}_2$ but with different constant curvature dependent only on $\rho$. The brane is cut off at $z=\epsilon$ or $\zeta=\epsilon\,\cosh\,\rho_*$. 
    }
    \label{fig:FGcoord}
\end{figure}

Near any asymptotic boundary, it is useful to introduce a 2D Fefferman-Graham coordinate system for $\Sigma$ so that
\begin{align}
    \d s_\Sigma^2=\frac{\d\zeta^2+\d\theta^2}{\zeta^2}.
\end{align}
From this, we can then also introduce the 3D Fefferman-Graham coordinate system via coordinate transformations
\begin{align}
    z = \frac{\zeta}{\cosh\, \rho},       \quad \tau =\zeta \tanh \,\rho.
\end{align}
This turns the metric \eqref{eq:coorwedge} into
\begin{align}
    \d s^2 = \frac{1}{z^2}\left(\d z^2+\d \tau^2+\d \theta^2\right).
\end{align}
Its relation to \eqref{eq:coorwedge} is illustrated in Figure~\ref{fig:FGcoord}.

In order to deal with asymptotic boundary terms and counterterms carefully, we need to introduce a cutoff. The cutoff surface is by definition $z=\epsilon$, represented by the dashed vertical line. The pink arc is where $\zeta=\epsilon$. The brane is cut off at $z=\epsilon$ or $\zeta=\epsilon\,\cosh\,\rho_*$.

In order to distinguish different asymptotic boundaries, we will label them with index $\a$, so we have coordinate patches $(\rho, \zeta_\a,\theta_\a)$ near each boundary $\a$. The label $\a$ will sometimes be suppressed. Note however that $\rho$ is a global coordinate, so it does not carry an $\alpha$ label.

We now evaluate the action \eqref{eq:action_EOW1}. Since we are in three dimensions, $R=-6$ and $\Lambda=-1$. Also, $K=2$ at the cutoff surface $z=\epsilon$. So the action simplifies to
\begin{align}
    I_{\rm wedge} 
    &= \frac{1}{4\pi G_N} \int_{\mathcal{M}} \sqrt{g}- \frac{1}{8\pi G_N} \int_Q \sqrt{h}\, \tanh\,\rho_*
    - \frac{1}{4\pi G_N} \int_{\CA} \sqrt{\gamma}.   
\end{align}
The bulk contributions can now be evaluated to the cutoff:
\begin{align}
    I_{\rm bulk}&=\frac{1}{4\pi G_N} \int_0^{\rho_*} d\rho\int_{\Sigma_\epsilon} {\cosh^2\rho}\,\sqrt{\sigma}
    -\sum_i\frac{1}{4\pi G_N} \int_0^{\rho_*} \d\rho\int_\epsilon^{\epsilon\,\cosh\,\rho}\d\zeta_i\int \d \theta_i\frac{\cosh^2\rho}{\zeta_i^2}.
\end{align}
The contribution from the brane up to the cutoff is given by:
\begin{align}
    I_{\rm brane}
    &=
    - \frac{1}{8\pi G_N} \tanh\,\rho_*\int_{\Sigma_\epsilon}\cosh^2\rho_*\, \sqrt{\sigma}
    +\sum_i\frac{1}{8\pi G_N} \tanh\,\rho_*\int_\epsilon^{\epsilon\,\cosh\,\rho_*}\d\zeta_i\int\d\theta_i\frac{\cosh^2\rho_*}{\zeta_i^2}.
\end{align}
The boundary term evaluated at the cutoff is given by:
\begin{align}
    I_{\rm asymp}&= \sum_i- \frac{1}{8\pi G_N} \int_0^{\rho_*} \left.\sqrt{g_{\rho\rho}\d\rho^2+g_{\zeta\zeta}\d\zeta^2}\right|_{z=\epsilon}\int\d\theta_i \frac{\cosh\rho}{\epsilon}.
\end{align}
Adding them up and taking the $\epsilon\to0$ limit, we obtain
\begin{align}
    I_{\rm wedge}
    &=-\frac{\rho_*}{16\pi G_N} \int_{\Sigma} \sqrt{\sigma}\,\mathcal{R}
    -\frac{\rho_*}{8\pi G_N}
    \int_{\partial\Sigma}\sqrt{\eta}\,K,
\end{align}
where we have used that on $\Sigma$, $\mathcal{R}=-2$ and on $\partial\Sigma$, $K=1$. Using the 2D Gauss-Bonnet theorem
\begin{align}
    \int_\Sigma \sqrt{h}\,\mathcal{R}+
    2\int_{\partial\Sigma}\sqrt{\eta}\,K=4\pi\chi,
\end{align}
we observe that this term is topological:
\begin{align}
\label{eq:wedgetopo}
    I_{\rm wedge}=-\frac{\rho_*}{4 G_N}\chi=-\chi\log g.
\end{align}
where $\chi$ is the Euler characteristic of $\Sigma$
\begin{align}
    \chi = 2 -2g-n-m,
\end{align}
where $g$ is the genus, $n$ is the number of asymptotic boundaries, and $m$ is the number of punctures which are topologically the same as boundary components.

Note that the topological nature of the EOW brane worked out here is consistent with the observation that there {\it does not} exist a Schwarzian mode on the 1D asymptotic boundary for the 3D wedge \cite{AKTW20}. This is also consistent with the observations that the effective 2D theory on the AdS$_2$ brane can be described by Liouville gravity (see e.g.~\cite{ST22,NSS24}).\footnote{Note that the AdS$_2$ brane cannot be described by JT gravity. This can also be seen from the fact that the brane admits on-shell configurations absent in JT gravity, such as thermal AdS$_2$. This fact becomes important in the construction of the random tensor model \cite{BBJNS23,JRW24} for BCFTs \cite{Jafferis:2025yxt}. Refer to \cite{NSS24} for an explanation why some works had expected the JT instead \cite{GKPRR22,DAZ22}, from an EFT point of view. }

\subsection{Negative tension}
\label{ssec:negtension}
As the tension is lowered, the wormhole shrinks in volume. This makes negative tensions more subtle to deal with than positive tensions. As the wormhole shrinks, the location of the brane moves inward while the rest of the geometry remains unchanged, which in particular includes the locations of the conical defects. It is then possible for a brane to encounter one or more conical defects \cite{KW22}. Let us define $T_{\rm crit}$ to be the tension when this first happens. Note that $T_{\rm crit}$ depends on the locations of the heavy bulk primary insertions.

As the tension is lowered further below $T_{\rm crit}$, we expect the configuration to look like (somewhat schematically)
\begin{align}
    \vcenter{\hbox{\includegraphics[height=3cm]{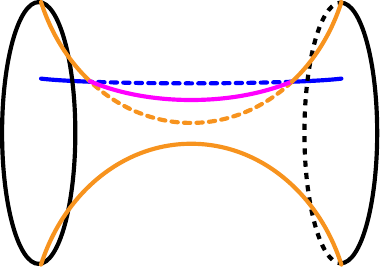}}}
\end{align}
which is a wormhole connecting two disks supported by some conical defects (only one shown). The blue line represents a conical defect. The dashed blue interval is where the conical defect would be if the tension were not as negative as $T_{\rm crit}$. The dashed orange line represents where the brane would be if the conical defect were absent. The pink line represents where the brane actually is. In an attempt to help the reader visualize it better, imagine that the brane folds around the defect, much like fabric draped over a clothesline. To see this more clearly, let us look at the evolution of the cross-section at the middle of the wormhole as the tension is lowered:
\begin{align}
    \vcenter{\hbox{\includegraphics[height=3cm]{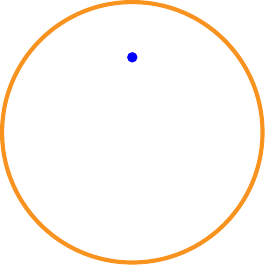}}}
    \longrightarrow
    \vcenter{\hbox{\includegraphics[height=3cm]{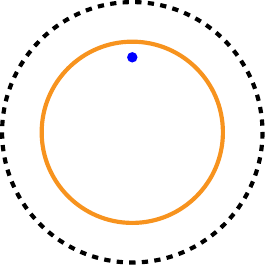}}}
    \longrightarrow
    \vcenter{\hbox{\includegraphics[height=3cm]{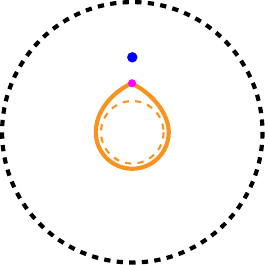}}}
\end{align}
where the blue dot represents the conical defect in the first two pictures and the would-be conical defect in the third picture, the pink dot represents a kink on the brane, and the dashed circle represents the would-be brane if there were no conical defect. The same pictures also represent cross-sections of the wormhole as we approach the middle from an asymptotic boundary. Notice that even for a given negative tension below $T_{\rm crit}$, some cross-sections of the brane (those closer to the asymptotic boundaries) are still perfectly circular and smooth. In other words, the kinks that appear this way do not connect boundary operators. We should not confuse these kinks with the ones appearing in the previous and next sections. 

We leave it as an open question to investigate whether such solutions always have the correct action, i.e., whether the action of such solutions agrees with the action of the corresponding geometries with positive tension analytically continued to below $T_{\rm crit}$. Evidence for this was presented in \cite{KW22}, but only for the case of global AdS.

\section{Most general wormholes}\label{sec:gen}

Without kinks, we showed in Section~\ref{sec:tension} that the tension of an EOW brane is ``topological" in the sense that we could replace the action of the tensionful brane with a tensionless one at the cost of inserting a topological term proportional to the Euler characteristic of the EOW brane. More precisely,
\begin{align}\label{eq:kinkfreereplace}
    I[\mathcal{M}] = I_{0}[\mathcal{M}_0] + I_{\rm topological}[Q_0],
\end{align}
where $I_{0}$ is the total action \eqref{eq:action_tot} with the tension $T$ set to zero, $\mathcal{M}_0$ is the on-shell manifold of $I_{0}$, and $Q_0$ is the EOW in $\mathcal{M}_0$. 

We emphasize that $\mathcal{M}$ and $\mathcal{M}_0$ are different manifolds, and in particular $\mathcal{M}_0$ is not a solution to the original action $I$. This equivalence is therefore only true at the level of the on-shell action. 

It is also useful to point out that
\begin{align}
    I_{\rm topological}[Q_0] = I_{\rm topological}[Q]
\end{align}
even though $Q$ and $Q_0$ have different intrinsic and extrinsic geometries. We will therefore sometimes abuse notation by writing $Q$ when it should technically be $Q_0$ when only the topology is concerned.

In the most general case, the EOW branes have tension, punctures, and kinks. Determining whether the 3D gravity theory in Section~\ref{sec:3dgrav} still exactly reproduces 2D BCFT results is a non-trivial task. In what follows, we present a conjecture for the most general case and provide some evidence for it. If the conjecture is true, the matching between the 3D gravity model and the 2D BCFT ensemble is exact. 

\subsection{Conjecture}
\label{ssec:conj}
We now conjecture that 
\begin{align}\label{eq:kinkfreereplacegen}
    I[\mathcal{M}] = I_{0}[\mathcal{M}_0] + I_{\rm topological}[Q],
\end{align}
where
\begin{align}\label{eq:topogen}
    I_{\rm topological} = -\sum_a \chi_a \log g_a + \frac{1}{2}\sum_{I} \chi_I(\log g_a+\log g_b).
\end{align}
Here, $\chi_a$ is the Euler characteristic of the (possibly punctured but not kinked) EOW brane segment $Q_a$ and $\chi_I$ is the 1D Euler characteristic of the kink $I$, which is by definition zero for a circle and one for a line. In the section term, for each kink $I$, $g_a$ and $g_b$ are the $g$-functions of the brane segments on either side of $I$, which are generally different. We should emphasize here that a puncture is not part of the brane $Q$, as defined in Section~\ref{ssec:alt}.

When there are no kinks, this action reproduces the results in Section~\ref{sec:tension}. When the $g_a$'s are all equal, i.e., all tensions are equal, the second term of the action \eqref{eq:topogen} simplifies and
\begin{align}\label{eq:topogen2}
    I_{\rm topological} &= -\sum_a \chi_a \log g + \sum_{I} \chi_I\log g\nn
    &=- \chi_Q \log g,
\end{align}
where $\chi_Q$ is the Euler characteristic of the entire surface $Q=(\sqcup_a Q_a)\sqcup (\sqcup_I \mathcal{K}_I)$. As an example, consider the wormhole \eqref{fig:1bk2bdworm}. It has two EOW branes with disk topology, each contributing $-\log g$ to the action, and two line kinks, each contributing $\log g$. The total topological action is zero. If we use \eqref{eq:topogen2}, we consider the whole brane $Q$, which has the topology of an annulus, so we get zero immediately. 

In higher dimensions, it is unlikely for a similar statement to hold because the geometry backreacts according to the locations of the dynamical objects such as the EOW brane in a highly sensitive way. In 3D Einstein gravity, the geometry is locally AdS${}_3$, so the backreaction is more under control. This rigidity of 3D gravity makes it conceivable for the conjecture to hold. 

From the ensemble perspective, the fact that this topological term reproduces the correct $g$-function dependence essentially follows from index counting. Here we will only present examples, and it is straightforward to apply it to any given manifold. A more systematic derivation can be found in \cite{Jafferis:2025yxt}. Some of the simplest examples are just the second moments for $B$ and $D$ given by \eqref{eq:B_second} and \eqref{eq:D_second}. As a result of bootstrap (and our choice of the normalization), they both scale as $g$ to the zeroth power (no dependence). This is exactly reproduced by the topological formula when applied to the corresponding bulk saddles \eqref{fig:B2worm} and \eqref{fig:D2worm}.  Another slightly more involved example would be the two-copy observable \eqref{eq:2copyeg1result}. As we can see, the first term has no dependence, which is reproduced by applying the topological formula to \eqref{fig:1bk2bdworm}. The second term is more interesting. It is a factorized term with contributions from two factorized manifolds, each looking like a manifold with a single disk brane with one puncture and one kink (combine \eqref{fig:punc} and \eqref{fig:kink}). Each kink contributes $\sqrt{g_ag_b}$, and the brane contributes no factors because the punctured disk has $\chi=0$. This reproduced the expected factor in the second term of \eqref{eq:2copyeg1result}. 

Let us consider a more complicated example drawn in \eqref{fig:mixedworm}. It computes the inner product between blocks with index structures
\begin{align}
    \sum_m C_{ijm} D_{mI}^{(a)}  \quad \text{and}\quad \sum_{M,N} B_{MNI}^{(aaa)} D_{iM}^{(a)} D_{jN}^{(a)}.
\end{align}
The first block only has bulk index contractions, so there is no associated factor; the second block has two boundary index contractions, namely the repeated indices $M$ and $N$. Using the conventions of \cite{NT22}, which we reviewed in Section~\ref{sec:BCFT_preliminaries}, each repeated lower boundary index must be contracted with an inverse ``metric" $\mathfrak{g}^{IJ}=\delta_{IJ}/\sqrt{g_ag_b}$ (for $I=J\in\mathcal{H}_{\rm open}^{(ab)}$), which gives a factor of $1/g_a$ in this case. The resulting expression should therefore be proportional to $1/g_a^2$. Applying the topological formula to \eqref{fig:mixedworm} exactly reproduces this dependence.

\subsection{Supporting case}\label{ssec:eg}

Even though the rigidity of 3D gravity makes it conceivable that some local action can be equivalent to a simpler and more topological one, it is not clear whether the action in Section~\ref{sec:3dgrav} is indeed a successful candidate. We now provide an example to support the conjecture that it is.

Consider the BCFT partition function on a disk with two boundary operators inserted on opposite points:
\begin{align}
    \vcenter{\hbox{\includegraphics[height=3cm]{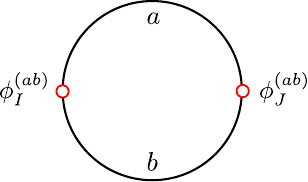}}}
\end{align}
which vanishes unless $I$ and $J$ are identical. For simplicity, we also take $g_a=g_b=g$. 

The simplest 3D manifold contributing to this partition function is given by a portion of the 3D Euclidean AdS geometry. We depict it as
\begin{align}\label{fig:diskbulk}
    \vcenter{\hbox{\includegraphics[height=5cm]{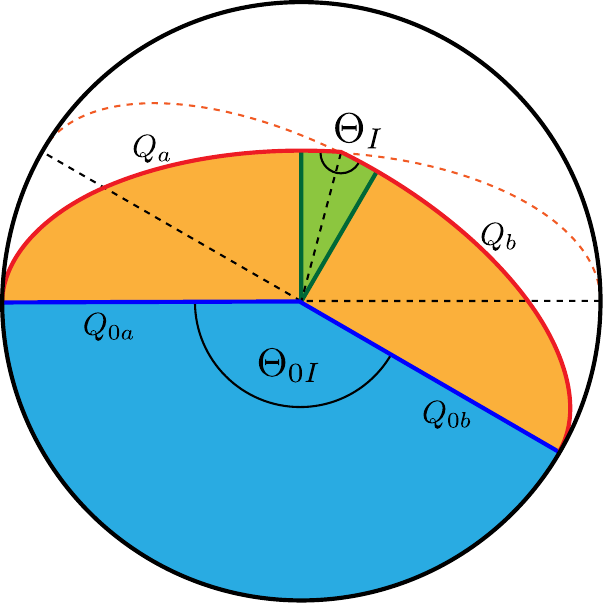}}}
\end{align}
which is a $\mathbb{Z}_2$-symmetric cross-section of Euclidean AdS${}_3$. The blue section is the wedge that would be dual to the disk two-point function if $g=1$, and $\Theta_{0I}$ is the kink angle fixed by the conformal weight of the operator $I$ via \eqref{eq:kinkweight}. Turning on the tension to a positive value makes the relevant manifold larger so that the orange and green sections are now included. To determine the shape of this manifold, we only need to know the change in the angle at which the tensionful brane intersects the asymptotic boundary as we have already seen in Figure~\ref{fig:FGcoord}. Doing this for both brane segments labeled by $a$ and $b$ entirely fixes the new location of the brane segments and in particular where they intersect, i.e., the location of the kink. The kink now has kink angle $\Theta_I$ which is different from $\Theta_{0I}$. As we commented after \eqref{eq:kinkweight}, knowing $\Theta_{0I}$ already allows us to determine the solution. We now understand why: $\Theta_{I}$ is determined geometrically from $\Theta_{0I}$ and the tensions.

From the BCFT point of view, we know that the disk two-point function is proportional to $g$. We now want to reproduce this from the action of the geometry shown in \eqref{fig:diskbulk}. More precisely,
\begin{align}
    Z_{\rm disk}(h_I,h_I)=g\,Z^0_{\rm disk}(h_I,h_I),
\end{align}
where $Z^0_{\rm disk}(h_I,h_I)$ is the disk two-point partition function if $g=1$, i.e., $T=0$. In 3D, this means that we want the action difference between $\mathcal{M}$ and $\mathcal{M}_0$ to be
\begin{align}\label{eq:want}
    I[\mathcal{M}] - I[\mathcal{M}_0] =-\log g = -\frac{1}{4G_N}\operatorname{arctanh} T.
\end{align}
However, notice that the orange wedges, which we now draw more clearly in 3D,
\begin{align}
    \vcenter{\hbox{\includegraphics[height=3cm]{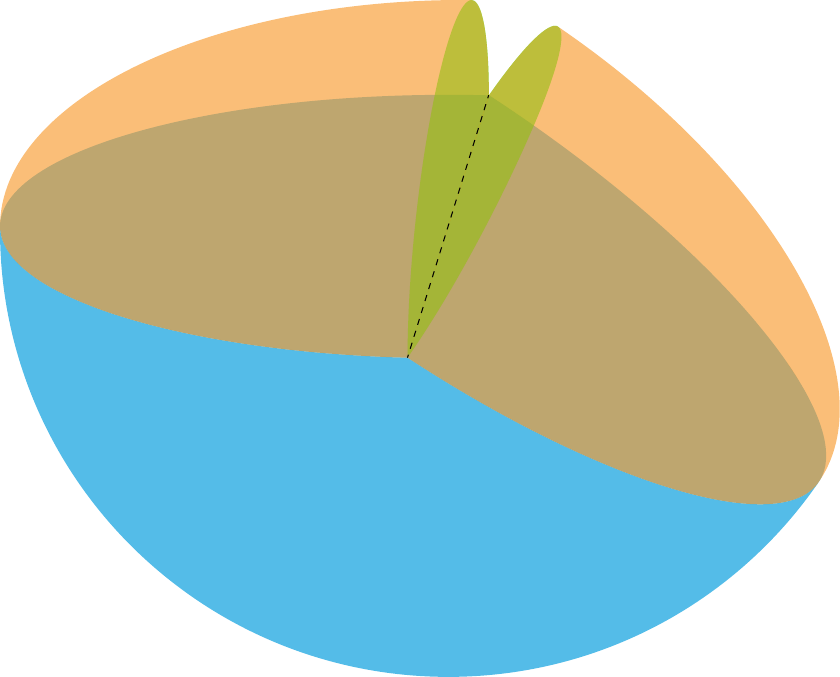}}} 
\end{align}
can be obtained by breaking the wedge appearing in \eqref{fig:diskwedge} like bread, we already know its action from Section~\ref{sec:g}, which is just $-\log g$. This means that in order for \eqref{eq:want} to hold, we need the green wedge shown in \eqref{fig:diskbulk} to have zero action. 

To show this, it is convenient to find a set of coordinates that simplifies the computation. In general, we can always write a locally AdS${}_{d+1}$ metric as
\begin{align}
    \d s^2_{\text{AdS}_{d+1}} = \d \rho^2 + \cosh^2\rho\, \d s^2_{\text{AdS}_{d}},
\end{align}
at least locally (say near $\rho=0$). Using this fact twice, we obtain the metric
\begin{align}\label{eq:doubleslicingcoor}
    \d s^2 = \d \rho^2 + \cosh^2\rho\left(\d \sigma^2+\cosh^2\sigma\,
    \frac{\d \xi^2}{\xi^2}\right).
\end{align}
Here we have picked the Fefferman-Graham coordinate for the metric of ``AdS${}_1$'', which is convenient for coordinate transformations we need to perform later, but the calculation works for the whole kink which has two asymptotic regions (not covered by a single Fefferman-Graham patch). 

This coordinate system covers half of the green wedge. Zooming in, the coordinate system looks like
\begin{align}\label{fig:diskkinkcoord}
    \vcenter{\hbox{\includegraphics[scale=1]{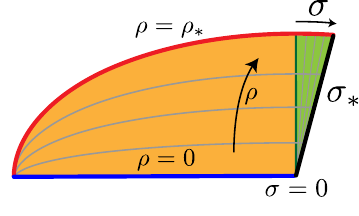}}} 
\end{align}
where the $\gamma$ coordinate is orthogonal to the page. The would-be tensionless brane $Q_{0a}$ is located at $\rho=0$ and $Q_a$ at $\rho=\rho_*$, related to the tension via \eqref{eq:Tandrhostar}. The vertical line separating the orange from the green is where $\sigma=0$, and $\sigma_*$ is the $\mathbb{Z}_2$-symmetric surface of the green wedge. 

The most technical part of the calculation is to determine $\sigma_{*}$ as a function of $\rho$. This can be achieved by going to a Fefferman-Graham coordinate system for the whole three-manifold near one of the boundary operator insertions. The required coordinate transformations are
\begin{alignat}{2}
    z &= \frac{\zeta}{\cosh\, \rho},       &\quad y &=\zeta
    \,\tanh \,\rho, \\
    \zeta &= \frac{\xi}{\cosh \,\sigma}, &\quad x &= \xi \,\tanh \,\sigma.
\end{alignat}
This turns the metric into
\begin{align}
    \d s^2 = \frac{1}{z^2}\left(\d z^2+\d x^2+\d y^2\right).
\end{align}
One of the boundary operators is located at $z=x=y=0$. In this coordinate system, we have already seen from Section~\ref{sec:tension} that each smooth section of the brane is a plane in $\mathbb{R}^3$ with $(x,y,z)$ being the (right-handed) Cartesian coordinates. We can also visualize the branes by demarcating their locations in the $(x,y)$ plane at different values of $z$:
\begin{align}
    \vcenter{\hbox{\includegraphics[height=4cm]{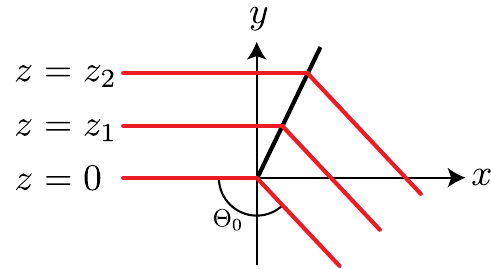}}} 
\end{align}
where $z_1>0$ and $z_2>z_1$ mark two constant-$z$ slices of the branes. The thick black line is the $\mathbb{Z}_2$-symmetric surface of the green wedge, and its angle from $y$ is $(\pi-\Theta_0)/2$, independent of $z$. The tensionless branes are not shown, but they coincide with the branes shown at $z=0$ and they are independent of $z$. 

Also notice that we did not label the angle $\Theta$ in this diagram because the angle should be defined in a local Cartesian coordinate system, i.e., they should be proper angles. We can identify $\Theta_0$ because it is the angle between the two tensionless branes whose locations are independent of $z$. In other words, the metric in the $(x,y)$ plane only changes by an overall scaling a function of $z$, so it does not change the angle. The same cannot be said about the angle between the tensionful branes (even at $z=0$). The relation between $\Theta_I$ appearing in \eqref{eq:actionkink} and $\Theta_{0I}$ is therefore a complicated expression. As we have just seen, however, we do not need to know the value of $\Theta_I$ in order to determine the solution as long as we know $\Theta_{0I}$ from \eqref{eq:kinkweight}. As we will now see, we can also determine the action without knowing the explicit relation between $\Theta_{I}$ and $\Theta_{0I}$.

To determine $\sigma_*$ as a function of $\rho$, we just need to find the function that describes the surface denoted by a black line in \eqref{fig:diskkinkcoord}. In $(x,y,z)$ coordinates, this is easy to write down: It is the surface defined by
\begin{align}
    x = y \tan \left(\frac{\pi-\Theta_0}{2}\right).
\end{align}
Translating into the coordinates \eqref{eq:doubleslicingcoor}, we obtain
\begin{align}
    \sinh\,\sigma_* = \tanh\,\rho \tan\left(\frac{\pi-\Theta_0}{2}\right).
\end{align}

Knowing the location of $\sigma_*$, we are ready to compute the action of the green wedge. The bulk action is given by
\begin{align}
    I_{\rm bulk}&=\frac{2}{4\pi G_N}\int \frac{\d\xi}{\xi} \int _0^{\rho_*} \d \rho\int_0^{\sigma_*(\rho)}\d\sigma\,\sqrt{g}
    \\
    &=\frac{1}{4\pi G_N}\sinh^2\rho_* \cot\frac{\Theta_0}{2}\int \sqrt{\gamma},
\end{align}
and the contribution from the brane is given by
\begin{align}
    I_{\rm brane}&=-\frac{2}{8\pi G_N}\left.\left[\int \frac{\d\xi}{\xi} \int_0^{\sigma_*(\rho)}\d\sigma\,\sqrt{h}\right]\right|_{\rho=\rho_*}\tanh\,\rho_*
    \\
    &=-\frac{1}{4\pi G_N}\sinh^2\rho_* \cot\frac{\Theta_0}{2}\int \sqrt{\gamma}.
\end{align}
The two contributions cancel. Notice that even though we have used the coordinate $\xi$ for the direction along the link at the beginning, it appears as a covariant expression, which is the proper length along the kink.  This concludes the example.

We also present an alternative calculation in the embedding space coordinates in Appendix~\ref{sec:embed}.

\section{Variants and remarks}
\label{sec:remarks}

\subsection{First moments and one-boundary wormholes}
\label{ssec:firstmom}

In pure 3D gravity with or without conical defects, odd moments of $C_{ijk}$ automatically vanish. From the bulk point of view, this is because conical defects cannot intersect and must end at asymptotic boundaries. A variant of the model could, for example, allow three defects to join at a point in the interior. There would then exist a saddle when the asymptotic boundary is a single three-punctured sphere. To match the gravity answer, the ensemble must now have $\overline{C_{ijk}}\ne0$, i.e., a non-zero mean. Higher moments will correspondingly change because of the existence of more complicated geometries with these triple joints. 

In our model, the punctures introduced in Section~\ref{sec:3dgrav} lead to a non-zero answer for $\overline{D_{i\id}^{(a)}}$ because of the saddle shown in \eqref{fig:sphere2pt}. This puncture is important for the match with the ensemble, as demonstrated by the example of \eqref{fig:sphere2pt} and the factorized term in \eqref{eq:2copyeg1result}. Along this line, we could also take a left-right $\mathbb{Z}_2$ quotient of the $C^2$ wormhole shown in \eqref{fig:C0}, which results in a ``one-boundary" wormhole\footnote{Although this geometrically resembles a half-wormhole, which can be important for restoring factorization in certain theories \cite{SSS21}, the one-sided wormhole here does not seem to probe anything finer than the average of the ensemble.}
\begin{align}
    \vcenter{\hbox{\includegraphics[height=2cm]{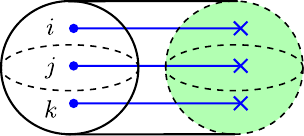}}}
\end{align}
where both the asymptotic boundary (left) and the finite EOW brane boundary are three-punctured spheres. These are higher-dimensional analogs of the JT EOW branes studied in \cite{GJK21}. This is not considered part of the solution space of the model in Section~\ref{sec:3dgrav}, but one can consider a variant of the model that allows this. This would be another way to introduce a non-vanishing $\overline{C_{ijk}}$. 

We want to emphasize that the puncture in our model which leads to $\overline{D_{i\id}^{(a)}}\ne0$ is less arbitrary. It is a consequence of a bootstrap argument when non-vacuum sub-threshold boundary operators are present \cite{Kusuki22}. A much simpler variant of our model was studied in \cite{KW22,Kusuki22} where the ensemble has $\overline{D_{i\id}^{(a)}}=0$ and contains no discrete list of boundary operators. The gravity model would correspondingly have no punctures or kinks. Our model can be considered a generalization of that model.

Another way to think of the non-vanishing nature of ${D_{i\id}^{(a)}}$ is to compare it with ${C_{ij\id}}$, which is just the normalization of the two-point function on the sphere and non-vanishing for $i=j$. The difference is that $C_{ii\id}=1$ is a constant (it only has a first moment) for all states, whereas $D^{(a)}_{i\id}$ has averaging for black hole states and no averaging for sub-threshold states.

\subsection{Higher moments and multi-boundary wormholes}

In Sections~\ref{sec:tensionless}, \ref{sec:tension}, and \ref{sec:gen}, we have kept the discussion general in terms of the number of asymptotic boundaries. In other words, the quadrupling, doubling, and wedge tricks, among others, all apply in the case of multiboundary wormholes. 

We have not given explicit expressions for higher moments because the explicit metrics for the wormholes and their actions are difficult to work with in the metric formulation of 3D gravity. In the formalism of Virasoro TQFT, multi-boundary wormholes can be worked out relatively easily (modulo numerical difficulties in evaluating integrals). 

Multi-boundary wormholes constitute an essential aspect of the ensemble interpretation of 3D gravity because they compute higher moments of the ensemble \cite{BdBL21}. Interestingly, higher-point crossing symmetries also yield expressions for universal asymptotics involving more factors of the OPE coefficients $C_{ijk}$ \cite{ABdBL21}. Using Virasoro TQFT, the relation between classical 3D gravity and the CFT ensemble is uplifted to finite $c$. An example for understanding the agreement between the gravity answer and the bootstrap answer is presented in Section~\ref{sec:threshold}. Multi-boundary wormholes and higher moments can be understood along those lines. 

In the context of BCFT, we are not aware of explicit computations of higher moments via bootstrap. As the BCFT bootstrap involves additional independent constraints \cite{Lewellen91}, there must be new results not attainable from the closed CFT bootstrap. Nevertheless, in our ensemble as well as that of \cite{CCHM22}, we have restricted ourselves to allowing only scalar bulk operators, a crucial ingredient in the quadrupling and halving tricks. In this case, we do not need the extra bootstrap conditions. In other words, the additional BCFT bootstrap conditions are needed to obtain universal asymptotic formulae associated to spinning bulk fields on bordered Riemann surfaces. It would be interesting to see this explicitly.

\subsection{BCFT Schlenker-Witten theorem}
\label{sec:SW}

As commented earlier in Section~\ref{ssec:firstmom}, one can engineer different ensembles, and the bulk theory would be different. Disallowing the conical defects in the model of \cite{CCHM22}, for example, would lead to an ensemble without the discrete list of heavy scalars below the black hole threshold. In this simpler setup, stronger results have been obtained, including for example what we refer to as the Schlenker-Witten theorem, which is (a special case of) one of the main results of \cite{SW22}.

Consider a CFT partition function on a closed Riemann surface of genus $g$. When a circle is pinched, it grows exponentially because of the negative energy of the vacuum. In gravity, as the path integral at large $c$ is given by $e^{-I}$, the exponential growth of the partition function corresponds to the action of the bulk manifold $\CM$ going to infinity in the pinching limit. It is shown in \cite{SW22} that this only happens if the circle being pinched is the boundary of a disk in $\CM$. If this holds for all the $3g-3$ independent circles, then it is called a \emph{sub-threshold amplitude}.

The Schlenker-Witten theorem states that for a Riemann surface $\Sigma$ that is part of the boundary of an on-shell manifold $\CM$, if all of the $3g-3$ non-intersecting, non-contractible, and homotopically independent circles of $\Sigma$ are compressible in $\CM$, then $\CM$ is a Schottky manifold and $\Sigma$ is the only boundary component of $\CM$.\footnote{They prove a stronger result where only $3g-5$ compressible circles are needed.}  

From the CFT point of view, the interpretation is that sub-threshold states do not exhibit averaging. Here, sub-threshold means just the vacuum, but adding light matter fields will not change the statement. Adding order-$c$ matter (such as conical defects), however, does \cite{SW22,CCHM22}.

A consequence of the conjectured formula \eqref{eq:kinkfreereplacegen} is that we can now use the quadrupling and halving tricks of Section~\ref{sec:tensionless} to study wormholes even in the most general case. However, we will consider a simpler setting where only EOW branes are allowed (no conical defects, no kinks, and no punctures). In this case, the formula \eqref{eq:kinkfreereplacegen} reduces to \eqref{eq:wedgetopo}, proved in Section~\ref{sec:tension}. This corresponds to an ensemble of BCFTs with only the vacuum and black hole states (plus light matter if one likes) in both the open and closed sectors. In this case, we can ask if a BCFT version of the Schlenker-Witten theorem holds. 

First of all, let us define the BCFT version of the sub-threshold amplitude. Using the quadrupling trick of Section~\ref{ssec:quadr}, we turn a BCFT partition function on a genus-$g$ Riemann surface with $n$ boundaries $\Sigma_{g,n}$ into a closed Riemann surface with genus $2g+n-1$, $\Sigma'_{2g+n-1,0}$. We define the BCFT amplitude to be a sub-threshold amplitude if the corresponding partition function computed on $\Sigma'_{2g+n-1,0}$ grows exponentially when any of the $6g+3n-6$ independent circles is pinched. 

As an example, consider $\Sigma_{0,3}$ with boundary conditions $a$, $b$, and $c$. The quadrupling trick gives a genus-two closed Riemann surface $\Sigma'_{2,0}$:
\begin{align}\label{fig:SWblue}
    \vcenter{\hbox{\includegraphics[height=3cm]{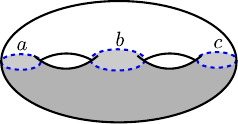}}}
\end{align}
where the shaded region is $\Sigma_{0,3}$, and the three blue circles are three independent circles, along which we can cut $\Sigma'$ into two pairs of pants. Alternatively, we can choose another set of three independent circles:
\begin{align}\label{fig:SWred}
    \vcenter{\hbox{\includegraphics[height=3cm]{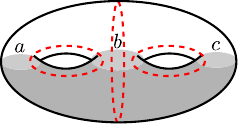}}}
\end{align}
where we can now cut along the red circles, which also gives two pairs of pants. 

From the perspective of $\Sigma'$, there is not a big difference, but they correspond to different ways of thinking about it in the original surface $\Sigma$. The first picture presents the decomposition of the partition function as
\begin{align}
    Z(\Sigma_{0,3}) = \sum_{i,j,k} D_{i\id}^{(a)}D_{j\id}^{(b)}D_{k\id}^{(c)}C_{ijk}\left|\mathcal{F}^{\rm (closed)}(P_i,P_j,P_k; m_1,m_2,m_3)\right|^2.
\end{align}
Note that $D_{i\id}^{(a)}$ vanishes for spinning fields.
The second picture represents
\begin{align}
    Z(\Sigma_{0,3}) =\sum_{I,J,K} B_{JII}^{(bba)}B_{JKK}^{(bbc)}\mathcal{F}^{\rm(open)}(P_I,P_I,P_J; \tilde{m}_1,\tilde{m}_2) \mathcal{F}^{\rm(open)}(P_K,P_K,P_J; \tilde{m}_3,\tilde{m}_2).
\end{align}
Focusing on $\Sigma_{0,3}$, the red lines are intervals that can be thought of as open state cuts. From the BCFT perspective (without performing the quadrupling trick), the definition of BCFT sub-threshold amplitudes therefore becomes the statement that the partition function on $\Sigma_{g,n}$ should grow exponentially when all state cuts are pinched, where each state cut is either open or closed. 

When an open state cut (an interval) is pinched (shrunk to a point), the corresponding circle in the doubled surface $\Sigma'$ is pinched. As shown in \cite{SW22}, the partition function grows exponentially only when this circle is the boundary of a disk in the bulk manifold $\CM'$. In the halved manifold, it becomes the statement that the union of the interval and the EOW brane must be the boundary of a (half) disk in $\CM$. 

We just gave an interlude that explains the interpretation in the original picture of $\Sigma$ and $\CM$, but for the proof of the BCFT Schlenker-Witten theorem, we can work entirely with $\Sigma'$ and $\CM'$. In $\CM'$, we can directly apply the original Schlenker-Witten theorem to conclude that there the only boundary of $\CM'$ is $\Sigma'$ if $Z(\Sigma')$ and therefore $Z(\Sigma)$ is a sub-threshold amplitude. 

This almost concludes the derivation of the BCFT Schlenker-Witten theorem. The only thing that we omitted is the fact that the quadrupling and halving tricks only work for zero-tension branes. However, using the results proven in Section~\ref{sec:tension}, adding tension only changes the action by a topological term, which is finite. As the proof of the Schlenker-Witten theorem is concerned with divergences, this finite term in the action does not actually change the conclusion. This concludes our argument that the BCFT version of the Schlenker-Witten theorem holds.

\subsection{The threshold is a thin line}
\label{sec:threshold}

There are two seemingly puzzling facts about the appearance of the $C_0$ function \eqref{eq:C0def}. The first emerges in the context of 2D bootstrap, and the second concerns 3D gravity. Both are tied to the observation that the function $C_0$ is rather universal, despite the qualitative difference in either the derivation or the geometric realization that depends on the weights of the objects.  In this sense, we say that there is a thin line between states above and below the black hole threshold.

It was noted in \cite{CCHM22} that the two issues are related. We will now make some comments regarding each of these two issues and hope that they can also shed light on their connections.

\subsubsection*{Light vs heavy in 2D bootstrap}

As reviewed in Section~\ref{sec:BCFTensemble}, a universal formula for OPE coefficients was obtained for $\overline{C^2_{ijk}}$ in \cite{CMMT19} using a purely 2D bootstrap approach. The formula was in fact derived as three independent formulae: when one, two, or all three of the states are heavy and averaged over while light ones (if any) are kept fixed. What is interesting is that the resulting formulae all look the same. 

In the first case, two of the states are light and fixed, while the third one is heavy and integrated over. The bootstrap method for deriving $\overline{C^2_{ijk}}$ in this case considers four points on the sphere. The four-point function can be computed in two ways, namely in the $s$-channel or in the $t$-channel. One then considers the kinematic limit where the identity operator dominates the $t$-channel. The corresponding conformal blocks can be represented pictorially as 2D pictures by
\begin{align}\label{fig:CLLH}
    \vcenter{\hbox{\includegraphics[height=3cm]{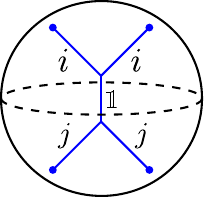}}}\quad = \quad 
    \sum_k~
    \vcenter{\hbox{\includegraphics[height=3cm]{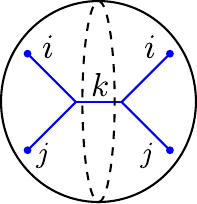}}}
\end{align}
where the blue dots represent operator insertions (or external states), the blue lines represent the propagation of states, and the dashed circles represent state cuts. In the first picture, the intermediate state is the identity; in the second picture, the propagating line represents a complete basis labeled by $k$. The idea here is that the $s$-channel involves the OPE coefficients $C_{ijk}^2$ while the $t$-channel involves only known functions. 

Following the logic of a 3D TQFT, which in this case is the Virasoro TQFT \cite{CEZ23}, we can understand the bootstrap result by considering the inner product of these two conformal blocks. The trick here is to embed the 2D diagrams as drawn in 3D and imagine each of the 2D spheres as the boundary of a 3-ball. Next, deform the blue lines into the interior of the 3-ball while keeping the dots anchored on the spheres. This gives the 3D TQFT partition function that prepares the TQFT state in the Hilbert space on the sphere. This follows from the important insight from the Chern-Simons/Weiss-Zumino-Witten correspondence \cite{Witten89} that the space of conformal blocks in the CFT is identified with the Hilbert space of the TQFT. Doing this for each of the diagrams above now gives us two 3-balls with a Wilson line network. Taking the inner product amounts to gluing them along the spheres, making sure the external operators, which are anchored on the spheres, are identified. A little mind exercise shows that the result is the following Wilson network embedded in $S^3$:
\begin{align}
    \vcenter{\hbox{\includegraphics[height=2.5cm]{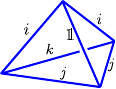}}}
\end{align}
which can be thought of as the edges of a pyramid. 

We will return to this picture soon, but let us now turn to the case of the bootstrap with two fixed light states and one averaged heavy state. One needs to now consider two identical operators on the torus \cite{CMMT19}. As before, the key idea is to express it in different channels. The two channels one needs to consider in this case are:
\begin{align}\label{fig:CLHH}
    \vcenter{\hbox{\includegraphics[height=2.5cm]{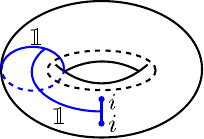}}}\quad = \quad \sum_{j,k}~
    \vcenter{\hbox{\includegraphics[height=2.5cm]{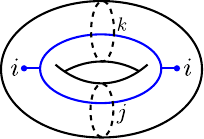}}}
\end{align}
where the first one is often referred to as the OPE channel and the second one as the necklace channel. The relevant kinematic limit here is where both internal states in the OPE channels are dominated by the identity. Again, one channel involves the OPE coefficients $C_{ijk}^2$ and the other one involves only known functions. 

To take the inner product, let us again consider it from the TQFT perspective. To obtain the 3D TQFT diagram corresponding to the first conformal block, we first embed the torus in $\mathbb{R}^3$ as drawn and then deform the blue skeleton into the interior of the solid torus. For the second picture, we do almost exactly the same thing, except that we want to embed it into $\mathbb{R}^3$ and then consider the exterior of the torus in $\mathbb{R}^3$ plus a point at infinity as the solid torus. Gluing them along the punctured tori gives the Wilson line network
\begin{align}
    \vcenter{\hbox{\includegraphics[height=3cm]{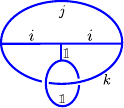}}}
\end{align}
in $S^3$. Ignoring the Wilson lines, the splitting of $S^3$ into two solid tori is a simple example of Heegaard splitting.

We can now understand the indifference between light and heavy states by noticing that both Wilson networks are the same once we remove the lines labeled by $\mathbbm{1}$. The reason we can remove those lines follows from the CFT choice of basis $C_{\mathbbm{1}ij}=\delta_{ij}$. To make the connection to $\overline{C^2}$, notice that upon removing a ball around the neighborhood of each trivalent Wilson junction, we simply get the two-boundary wormhole \eqref{fig:C0}, which has the topology of a three-punctured sphere times the interval. This is known to reproduce the Gaussian statistics of the OPE coefficients \cite{BdB20}. 

So far, we have obtained two different Wilson line networks in $S^3$ by considering the bootstrap procedures for obtaining $\overline{C_{\rm LLH}^2}$ and $\overline{C_{\rm LHH}^2}$ respectively, where L and H denote light and heavy. There is a similar procedure for $\overline{C_{\rm HHH}^2}$ which would require consideration of the genus-2 partition function \cite{CMMT19}, and doing so would give us another Wilson line network in $S^3$. The story is similar.

The lesson we want to draw from this observation is that Virasoro TQFT provides a realization (or at least a way of visualization) that unifies the different ways of deriving the same formula depending on whether states are below or above the black hole threshold. In fact, the Wilson lines in Virasoro TQFT can either have real or imaginary Liouville momentum, so it can be either above or below the threshold.

The story is pretty much the same for BCFTs. The BCFT bootstrap for computing the universal OPE statistics of $\overline{B_{IJK}^2}$ also divides into three cases: light-light-heavy, light-heavy-heavy, and heavy-heavy-heavy \cite{Kusuki21,NT22}. For $\overline{B_{\rm LLH}^2}$, one considers the disk with four boundary operators, for $\overline{B_{\rm LHH}^2}$, one considers the annulus with two boundary operators, and for $\overline{B_{\rm HHH}^2}$, one considers the partition on the sphere with three circle boundaries. We immediately notice that all of the cases are the $\mathbb{Z}^2$ quotient versions of the bootstrap for $\overline{C^2_{ijk}}$, so the 3D diagrammatics essentially extends to this case. 

The computation for the universal OPE statistics of $\overline{D_{iI}^2}$ is also similar \cite{Kusuki21,NT22}. For $\overline{D_{\rm HL}^2}$, consider the bulk two-point function on the disk, for $\overline{D_{\rm LH}^2}$, consider the boundary two-point function on the annulus, and for $\overline{D_{\rm HH}^2}$, consider the partition function on the torus with a circular boundary (hole). These are again $\mathbb{Z}_2$ quotients of the corresponding objects studied in the bootstrap for $\overline{C^2_{ijk}}$, except that they are of different $\mathbb{Z}_2$ groups.

More generally, the idea works for higher moments of $C$, $B$ and $D$, and also mixed moments such as $\overline{C_{ijk}B_{IJK}^{(aaa)}D^{(a)}_{iI}D^{(a)}_{iJ}D^{(a)}_{kK}}$ (see Section~\ref{ssec:mix}), though both the bootstrap method for deriving them and the corresponding gravity calculations involve more complicated diagrams.

\subsubsection*{Crossing the threshold in 3D gravity}\label{eq:cross}

Let us now discuss a similar problem in 3D gravity. It is explained in \cite{CCHM22} that analytically continuing the conformal dimension of the operators corresponding to the conical defects above the black hole threshold (and integrating over the black hole continuum) gives the correct partition function for a new geometry obtained by replacing the defect with a tube. The topology of the wormhole and even the number of boundaries change abruptly, but the action changes in a predictable manner. 

For example, analytically continuing $P_k$ and then $P_j$ in $C_0(ijk)$ corresponds to 
\begin{align}\label{fig:oneboundary4point}
    \vcenter{\hbox{\includegraphics[height=2cm]{figs/C0.pdf}}}
    \longrightarrow     \vcenter{\hbox{\includegraphics[height=2cm]{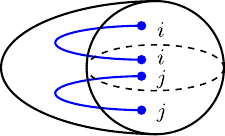}}}
    \longrightarrow  \vcenter{\hbox{\includegraphics[height=2cm]{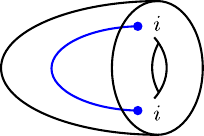}}}
\end{align}
where the first diagram has two sphere boundaries, the second has one, and the third has a torus boundary. In terms of topology, analytically continuing the conformal dimension corresponding to the conical defect above the black hole threshold amounts to removing a tubular neighborhood of the conical defect (which connects two asymptotic boundaries) and identifying the annulus (which is the boundary of the tube) as a part of the new asymptotic boundary. In each step, the conformal weight of the conical defect becomes a new moduli parameter of the new boundary geometry.

From a TQFT perspective, the second diagram in \eqref{fig:oneboundary4point} is exactly the same as the $t$-channel diagram in \eqref{fig:CLLH} if we deform the skeleton lines into the interior and remove the identity line, and similarly the third diagram in \eqref{fig:oneboundary4point} is exactly the same as the OPE channel diagram in \eqref{fig:CLHH}. In this sense, the two issues are essentially the same.

Let us now comment on the tunneling process as viewed from the TQFT perspective. For this purpose, it is useful to recall that the Cardy density is defined via the modular $S$-transformation. So integrating over heavy states amounts to
\begin{align}
    \int \d P\d \bar{P} \,|\rho_0(P)|^2 = \int \d P\d \bar{P} \,|\mathbb{S}_{\id P}[\id]|^2, 
\end{align}
where $\mathbb{S}_{jk}[i]$ is the modular $S$-matrix that relates the one-point function of bulk operator $i$ on the torus to the dual channel. In Virasoro TQFT, for a Wilson loop with momentum $P$, performing this integral is equivalent to performing the toroidal surgery \cite{JRW24}: remove the toroidal neighborhood of the Wilson loop and glue back a solid torus with the two boundary circles swapped. For Wilson lines extending between points on asymptotic boundaries (or just one asymptotic boundary), one should perform ``half'' of the toroidal surgery, which amounts to replacing the Wilson line with a tube connecting the boundaries. This is also explained in \cite{JRW24}.

For the model in Section~\ref{sec:3dgrav}, the analog of the toroidal surgery is an ``annular'' surgery, where we remove an annular neighborhood of a boundary Wilson loop (the finite-$c$ version of the kink) and glue back a slab (an object with the topology of a disk times the interval, whose boundary is composed of an annulus and two disk EOW branes with generally different boundary condition labels). For a boundary Wilson line extending from a BCFT boundary operator to another, we perform ``half'' of the annular surgery, which amounts to digging a trench (a strip neighborhood) around the line and identifying it as part of the BCFT manifold. Another way to visualize the annular surgery and half-annular surgeries is to use the halving trick explained in Section~\ref{ssec:halve} on the toroidal and half-toroidal surgeries. These can be made more precise in an open-closed extension of Virasoro TQFT \cite{Jafferis:2025yxt}.

\section{Discussion}
\label{sec:disc}

In this work, we constructed and studied wormholes in a 3D gravity model incorporating conical defects, EOW branes, kinks, and punctures, thereby extending the paradigm of ensemble averaging in AdS${}_3$/CFT${}_2$ to BCFTs. Classical wormholes with these ingredients provide a bulk realization of statistical moments of BCFT data and reproduce the formulae showing up in the universal asymptotics of 2D conformal bootstrap.

Our key findings are the following. We first presented a no-go statement prohibiting the existence of certain wormholes, from which we conclude that certain objects do not exhibit ensemble averaging. This includes in particular the $g$-functions and the OPE coefficients $D_{i\id}^{(a)}$ for sub-threshold scalars $\phi_i$. We then studied wormholes with tensionless branes. In this case, we outlined how to obtain wormholes with EOW branes and kinks from ones without them using what we call the quadrupling trick followed by the halving trick. For wormholes with tensionful branes, we first showed that introducing the brane tension modifies the geometry through the addition or deletion of wedge-like regions. For branes without kinks, we proved that the change of the action as a function of tension depends only on the topology of the brane, leading to a topological relation. For branes with kinks, we conjectured a more general topological relation involving Euler characteristics of both the branes and kinks and provided explicit computations supporting it. Throughout this paper, we have emphasized the important role of $g$-functions. The topological nature of the branes permits an agreement of the $g$-function dependence between 3D gravity and 2D BCFT ensemble. A BCFT analog of the Schlenker–Witten theorem also naturally emerges as a consequence of this topological structure. We also outlined the role of the punctures in contributing to consistency of the ensemble.

Let us now move on to some open questions and future directions. 

As we did not prove the conjecture about the topological action in the most general case, an immediate follow-up work would be to prove it or test it in more complicated examples. Relatedly, another question mentioned in the main text is the case of negative tensions. Investigating if the actions remain analytic in the brane tension as the brane touches the conical defects can be potentially interesting. We also did not systematically derive a variational principle for the action of EOW branes at the punctures, though we have control over their on-shell values as part of the construction. It would be interesting to study its variational principle.

The AdS/BCFT duality has also been proposed for higher dimensions. However, as gravity has local degrees of freedom in higher dimensions, unlike 3D pure gravity, it is perhaps unreasonable to expect that the brane action is equivalent to a topological one. On the other hand, it is not clear at all whether there is a dual ensemble description of gravity in higher dimensions. Even if there is one, the exact matching of $g$-functions would seem a difficult task to accomplish, not to mention that wormholes, especially multi-boundary ones, are already much harder to construct even in the case without boundaries.

In the formalism of \cite{CCHM22} and in ours, heavy operators above the black hole threshold are integrated over. It would be interesting to study single instances of above-threshold operators (huge operators). With huge operator insertions at the asymptotic boundaries, the bulk geometry has large backreactions. The resulting geometries are known as ``spacetime bananas" \cite{AAMV23a,AAMV23b}. Spacetime bananas with EOW branes have also been studied \cite{TLO24}. Formulating a duality between 3D gravity with multiboundary wormholes containing bananas and an ensemble of CFTs or BCFTs is potentially promising. It might also be interesting to apply the techniques of \cite{CHMN23} to our formalism or vice versa. For example, one might expect boundary operator insertions to correspond to sharp corners in Liouville CFT.

Given that we have approached BCFTs primarily as extensions of closed CFTs, it might appear that the discussion presented in this work is irrelevant to those concerned solely with closed CFTs. However, BCFTs can also be a useful tool for the study of closed CFTs themselves, offering insights that may not be readily accessible within the purely closed framework. For example, BCFT techniques are used to compute entanglement and Renyi entropies in a CFT with no borders \cite{OT14}. Also, a CFT can be studied by triangulation techniques, which decompose the partition function on a closed Riemann surface into BCFT ones plus some boundary conditions \cite{CHJL24,HJ24,BHJL24}. It would be interesting to establish a connection between our work and theirs.

Throughout this work, we have restricted ourselves to BCFTs on orientable surfaces. It would be natural to extend this to CFTs defined on non-orientable Riemann surfaces, i.e., those with crosscaps. Different holographic objects have been proposed to account for them when the CFT lives on a single closed Riemann surface \cite{MR16,Wei24}. First, it would be interesting to study these different bulk models in the context of CFT ensembles. Moreover, including both borders and crosscaps in the CFT can potentially lead to interesting interplays between different bulk defects.

Supersymmetry is another direction for generalization. For $\mathcal{N}=1$, bootstrap works similarly, with crossing kernels replaced correspondingly \cite{Hadasz07,CH08,CHJ11,PSS13}. At finite $c$, the crossing kernels can be used to explicitly construct the super-Virasoro TQFT in 3D \cite{BGNP24}. In the absence of sub-threshold states (except the vacuum), the large-$c$ limit yields $\mathcal{N}=1$ supergravity. It seems an interesting open question to understand the supersymmetric analogs of conical defects, kinks, punctures, etc., which are necessary for studying an ensemble that includes sub-threshold states.

In AdS$_3$ gravity, there are one-boundary wormholes with handles and conical defects going through them, as shown in \cite{CCHM22}. When the handles are ``integrated out", they induce random bilocal couplings in the bulk. This is related to the Coleman-Giddings-Strominger mechanism \cite{Coleman88,GS87,GS88}. For wormholes with EOW branes, it is plausible that such a mechanism exists for kinks as well, so it induces a random coupling on the brane. The details of this remain to be explored.

We hope this work paves the way for further explorations into the rich interplay between geometry, topology, conformal bootstrap, and defects.

\section*{Acknowledgements}
It is a pleasure to thank Jeevan Chandra, Scott Collier, Tom Hartman, Daniel Jafferis, Yikun Jiang, Don Marolf, Henry Maxfield, Liza Rozenberg, Julian Sonner, Douglas Stanford, Cynthia Yan, and Mengyang Zhang for interesting discussions. 
Diandian Wang acknowledges support by NSF grant PHY-2207659. Zhencheng Wang is supported by the DOE award number DE-SC0015655. Zixia Wei is supported by the Society of Fellows at Harvard University. 

\appendix

\section{Embedding space calculation of the disk two-point function}\label{sec:embed}
In this appendix, we give an alternative derivation of the example in Section~\ref{ssec:eg}, using the method of embedding space. 

The Euclidean AdS$_3$ geometry can be embedded in $\mathbb{R}^{3,1}$ as the hyperboloid
\begin{align}
-T^2+X^2+Y^2+Z^2=-1,
\end{align}
where $T$ is the time coordinate, and $X$, $Y$, and $Z$ are spatial coordinates. The EOW branes are codimension-one surfaces in Euclidean AdS${}_3$ such that $K_{\mu \nu} = T h_{\mu \nu}$ (which can be derived from \eqref{eq:braneEOM}). A surface whose extrinsic curvature is proportional to its induced metric is sometimes called a ``totally umbilical surface''. In the embedding space, such a surface (in our case, the brane) can be represented by the intersection of a plane $aT+bX+cY+dZ+e=0$ and the hyperboloid. As a special case, when $K_{\mu\nu}=0$ (corresponding to the tensionless brane) the surface is represented as the intersection of a plane passing through the origin $aT+bX+cY+dZ=0$ and the hyperboloid.

Consider now the example \eqref{fig:diskbulk}, and we want to compute the action of the green wedge.  Without loss of generality, let $X=0$ be the plane that intersects the hyperboloid at $Q_{0a}$. Then use constant-$X$ slices to foliate the hyperboloid, i.e., $X$ serves as a good coordinate in the intrinsic geometry of AdS$_3$. Each constant $X$ surface corresponds to a brane with a constant tension determined by the value of $X$. Without loss of generality, using the symmetries of the embedding space, we can choose $W\equiv X\sin\varphi  + Y\cos \varphi  =0$ to be the plane that intersects the hyperboloid at $Q_{0b}$. Constant-$W$ slices give another foliation of AdS$_3$. It can be shown that $\varphi$ is related to $\Theta_0$ defined in \eqref{fig:diskbulk} through $\sec\varphi = \cot \frac{\Theta_0}{2}$. Using $X$ and $W$, the AdS$_3$ metric can then be rewritten as
\begin{equation}
    \begin{aligned}
    \d s^2&=- \d T^2+\d X^2+\d Y^2+\d Z^2 \\
       &=\sec^2\varphi \,(\d X^2+\d W^2-2\d X \d W \sin\varphi)+
       \frac{\d z^2}{z^2}(1+\sec^2\varphi (X^2+W^2-2WX\sin\varphi))\\&\quad+\frac{\sec^2\varphi}{z}\left(2(X-W\sin\varphi)\d X\d z+2(W-X\sin\varphi)\d W\d z\right), 
    \end{aligned}
\end{equation}
where we have chosen the remaining one coordinate of AdS$_3$ to be the same as the $z$-coordinate in the Poincare coordinate system (but other choices are equally valid). It is related to $T$ and $Z$ via
\begin{align}
    T=\frac{1+t^2+x^2+z^2}{2z},\qquad Z=\frac{-1+t^2+x^2+z^2}{2z}.
\end{align}
It can be computed that each constant $X$ surface has $K= \frac{X}{\sqrt{1+X^2}}$ (and similarly for $W$). So when we tune the brane tension from $0$ to $T=\tanh \,\rho_*$, the corresponding $X$ (or $W$) coordinate ranges from $0$ to $\sinh\,\rho_*$. Then for the wedge, the bulk term evaluates to
\begin{equation}
\begin{aligned}
    I_{\rm bulk}&= \frac{1}{16\pi G_N}\int \d^3x \sqrt{g}\,(R+2)\\
    &=-\frac{\sec\varphi }{4\pi G_N} \int_0^{\sinh\,\rho_*}  \d X \int_0^{\sinh\,\rho_*} \d W \int \frac{\d z}{z}=-\frac{\sec\varphi \,\sinh^2 \rho_*}{4\pi G}\int  \frac{\d z}{z},
\end{aligned}
\end{equation}
where we used $R+2=-4$ and $\sqrt{g}=\sec\varphi /z$. The contributions from the boundary terms at the branes are computed as
\begin{equation}
\begin{aligned}
    I_{\rm brane}&=2\times \frac{1}{8\pi G_N} \int \d^2x \sqrt{h} (K-T)\\
    &=\frac{T\sec\varphi}{4\pi G_N} \int_0^{\sinh\,\rho_*} \d X \int \d z  \frac{\sqrt{1+X_*^2}}{z}= \frac{\sec\varphi\,\sinh^2\rho_*}{4\pi G} \int  \frac{\d z}{z},
\end{aligned}
\end{equation}
where the factor of $2$ comes from the two pieces of branes being accounted for, namely $Q_a$ and $Q_b$ (having the same tension in this case), and we also used $K=2T=2\,\tanh\, \rho_*$, $\sqrt{h}=\sec\varphi{\sqrt{1+X_*^2}}/{z}$. 

As we can see, the two parts of the action cancel out, so the green wedge in \eqref{fig:diskbulk} does not contribute to the action. This agrees with the result in the main text calculated using a different method.

\newpage
\bibliographystyle{jhep}
\bibliography{wormhole}

\end{document}